\RequirePackage{fix-cm}
\documentclass[twocolumn, epjc3, comma, sort&compress, natbib]{svjour3}
\bibpunct{[}{]}{,}{n}{}{,} % to get "[numbered]" references from natbib

\journalname{Eur. Phys. J. A}

\usepackage{latexsym}
\usepackage{amsmath}
\usepackage{amssymb}
\usepackage{amsfonts}
\usepackage{CJKutf8}

\usepackage[mathscr,scaled=1.15]{urwchancal}
\DeclareFontFamily{OT1}{pzc}{}
\DeclareFontShape{OT1}{pzc}{m}{it}%
{<-> s * [1.15] pzcmi7t}{}
\DeclareMathAlphabet{\mathpzc}{OT1}{pzc}{m}{it}

\usepackage{color}

\usepackage{supertabular}
\usepackage{placeins}
\usepackage{epsfig}
\usepackage{graphicx}

\definecolor{purple}{rgb}{0.5,0,0.5}
\definecolor{blue}{rgb}{0.0,0,0.9}
\definecolor{prdblue}{rgb}{0.133,0.118,0.498}
\usepackage[colorlinks=true, pdfstartview=FitV, linkcolor=prdblue, citecolor= prdblue, urlcolor=prdblue]{hyperref}

\begin{document}
\begin{CJK*}{UTF8}{gbsn}
%\begin{CJK*}{UTF8}{song}

\title{$\,$\\[-6ex]\hspace*{\fill}{\normalsize{\sf\emph{Preprint no}.\
NJU-INP 116/26}}\\[1ex]
Contact interaction treatment of the nucleon Faddeev equation}

\author{
     Xin-Yu Bai (白芯瑜)\thanksref{NJU,INP}%
       $^{,\href{https://orcid.org/0009-0007-4429-103X}{\textcolor[rgb]{0.00,1.00,0.00}{\sf ID}}}$
\and
Ya Lu (陆亚)\thanksref{NJT}%
       $\,^{\href{https://orcid.org/0000-0002-0262-1287}{\textcolor[rgb]{0.00,1.00,0.00}{\sf ID}}}$
\and
Zhao-Qian\ Yao (姚照千)\thanksref{HZDR}%
       $^{,\href{https://orcid.org/0000-0002-9621-6994}{\textcolor[rgb]{0.00,1.00,0.00}{\sf ID}},}$
%\email{z.yao@hzdr.de}
\and
     \\Craig D. Roberts\thanksref{NJU,INP}%
       $^{,\href{https://orcid.org/0000-0002-2937-1361}{\textcolor[rgb]{0.00,1.00,0.00}{\sf ID}}}$
\and
Sebastian M.\ Schmidt\thanksref{HZDR,TUD}%
    $\,^{\href{https://orcid.org/0000-0002-8947-1532}{\textcolor[rgb]{0.00,1.00,0.00}{\sf ID}},}$}
%\email[]{s.schmidt@hzdr.de}

\authorrunning{Xin-Yu Bai \emph{et al}.} % if too long for running head

\institute{School of Physics, \href{https://ror.org/01rxvg760}{Nanjing University}, Nanjing, Jiangsu 210093, China \label{NJU}
\and
Institute for Nonperturbative Physics, \href{https://ror.org/01rxvg760}{Nanjing University}, Nanjing, Jiangsu 210093, China \label{INP}
\and
Department of Physics, \href{https://ror.org/03sd35x91}{Nanjing Tech University}, Nanjing 211816, China \label{NJT}
\and
\href{https://ror.org/01zy2cs03}{Helmholtz-Zentrum Dresden-Rossendorf}, Bautzner Landstra{\ss}e 400, D-01328 Dresden, Germany \label{HZDR}
\and
\href{https://ror.org/042aqky30}{Technische Universit\"at Dresden}, D-01062 Dresden, Germany \label{TUD}
\\[1ex]
Email:
\href{mailto:z.yao@hzdr.de}{luya@njtech.edu.cn} (YL);
\href{mailto:z.yao@hzdr.de}{z.yao@hzdr.de} (ZQY);
\href{mailto:cdroberts@nju.edu.cn}{cdroberts@nju.edu.cn} (CDR)
%\href{mailto:jose.rodriguez@dfaie.uhu.es}{jose.rodriguez@dfaie.uhu.es} (JRQ)
            }

\date{2026 June 01} % 
%\date{2026 May 28} % revision submitted
%\date{2026 May 25} % revision begun
%\date{2026 February 16} % submitted
%\date{2026 February 02}
%\date{2026 January 18} -- begun

\maketitle

\end{CJK*}

\begin{abstract}
Working with a symmetry-pre\-ser\-ving treatment of a vector\,$\otimes$\,vector contact interaction (SCI), a largely algebraic three-body Faddeev equation treatment of the nucleon bound state problem is introduced and used to deliver results for all nucleon charge and magnetisation distributions and their flavour separation.  A strength of the SCI treatment is that it provides for a transparent understanding of this three-body approach to developing predictions for bar\-yon observables.  Comparisons of SCI results with predictions obtained in realistic-interaction Faddeev equation studies reveal the sensitivities of a given observable to the pointwise behaviour of the quark-quark interaction and phenomena associated with the emergence of hadron mass.
\end{abstract}

%\begin{keyword}
%continuum Schwinger function methods \sep
%Dyson-Schwinger equations \sep
%electromagnetic form factors \sep
%emergence of mass \sep
%nucleons - neutrons and protons \sep
%nonperturbative quantum field theory \sep
%quantum chromodynamics
%\end{keyword}

%%%%%%%%%%%%%%%%%%%%%%%%%%%%%%%%%%%%%%%%%%%%%%%%%%%%%%%%%%%%%%%%%%%%%%%%%%%%%%%%
%%%%%%%%%%%%%%%%%%%%%%%%%%%%%%%%%%%%%%%%%%%%%%%%%%%%%%%%%%%%%%%%%%%%%%%%%%%%%%%%

\section{Introduction}
%%%\label{secintro}
%\noindent\emph{1.\,Introduction}\,---\,%
The nucleon appears as a pole in the six-point quark Schwinger function, \emph{viz}.\ interactions produced by quantum chromodynamics (QCD) generate an isolated colour-singlet pole contribution to the three light-quark $\to$  three light-quark scattering matrix.
The residue at this pole is the nucleon's Poincar\'e-covariant Faddeev wave function.
As is widely known and we now explain, one may develop an approximation to this wave function by solving what may be called a three-body Faddeev equation \cite{Kievsky:2019}.

The first tractable approach to this formulation of the nucleon bound-state problem was introduced in Refs. \cite{Cahill:1988dx, Reinhardt:1989rw, Efimov:1990uz}, which exploited the pairing proclivity of fermions to simplify the bound-state equation.
The equation that emerges describes dressed quarks and fully-interacting quark-quark (diquark) correlations built therefrom, which bind together into a baryon, at least in part, because of the continual exchange of roles between the spectator and diquark-participant quarks.
Many efficacious studies of nucleon properties have employed this $q(qq)$ scheme \cite{Barabanov:2020jvn, Ding:2022ows, Cheng:2023kmt, Chen:2023zhh, Yu:2024ovn, Eichmann:2025tzm, Yu:2025fer, Cheng:2025yij}.

Notwithstanding the wide-ranging success of the $q(qq)$ approach, today, with high-performance computing resources readily available, it is becoming more common to tackle the three-body ($3$-body) Faddeev equation directly.  The first such study was reported in Ref.\,\cite{Eichmann:2009qa}.
It was quickly followed by $3$-body calculations of the masses of other octet baryons \cite{Sanchis-Alepuz:2011egq} and nucleon electromagnetic and axial form factors \cite{Eichmann:2011vu, Eichmann:2011pv}.
More recent studies have delivered, \emph{e.g}., predictions for the spectrum of
light- and heavy-baryons \cite{Qin:2018dqp, Qin:2019hgk},
updated results for nucleon electromagnetic form factors \cite{Yao:2024uej}
and predictions for nucleon gravitational form factors \cite{Yao:2024ixu}.
It is worth noting that, wherever comparisons have been made, $3$-body and $q(qq)$ results are largely in agreement \cite{Eichmann:2016yit, Cheng:2025yij, Xu:2025ups}.

Whether the $q(qq)$ scheme or the $3$-body approach is used, especially the latter, numerical analyses on a fairly large scale are required.  Similar statements are true for state-of-the-art analyses of meson problems using continuum Schwinger function methods (CSMs) for QCD.
Recognising this, a symmetry-pre\-ser\-ving treatment of a vector\,$\otimes$\,vector contact interaction (SCI) was introduced in Ref.\,\cite{Gutierrez-Guerrero:2010waf} with the goal of providing a tool that could widely be employed to develop insights into hadron observables and draw baselines for more sophisticated studies.
For instance, comparisons between SCI results and predictions obtained using QCD-connected interactions \cite{Qin:2011dd, Binosi:2014aea} serve to highlight a given observable's sensitivity to the pointwise behaviour of the quark-quark interaction and phenomena associated with the emergence of hadron mass (EHM) \cite{Roberts:2021nhw, Binosi:2022djx, Ding:2022ows, Roberts:2022rxm, Raya:2024ejx, Ferreira:2023fva, Achenbach:2025kfx}.

Since that first study, the SCI has been refined.  It is not a precision tool, but the modern formulation has many merits, such as:
algebraic simplicity;
simultaneous applicability to a large array of systems and processes;
potential for revealing insights that link and explain many phenomena;
and service as a tool for checking the viability of algorithms used in calculations that depend upon high performance computing.
Modern SCI applications are typically parameter-free and numerous benchmarking predictions are available for a wide range of phenomena involving mesons
\cite{Roberts:2011wy, Chen:2012txa, Serna:2017nlr, Zhang:2020ecj, Xing:2022sor, Cheng:2024gyv, Sultan:2024hep, Xing:2025eip, Gutierrez-Guerrero:2019uwa, Chen:2024emt, Gutierrez-Guerrero:2026rsb}
and the $q(qq)$ picture of baryons
\cite{Gutierrez-Guerrero:2019uwa, Chen:2024emt, Wilson:2011aa, Segovia:2013rca, Xu:2015kta, Yin:2019bxe, Raya:2021pyr, Cheng:2022jxe, Yu:2025fer}.
Given its established utility in $q(qq)$ studies of baryons, herein we introduce a SCI treatment of the nucleon $3$-body Faddeev equation and the associated calculation of nucleon electromagnetic form factors.

%Now SCI for 3-body ... not competitive in reality with realistic studies but ease of application means it can be applied to many problems as an exploratory tool to lay ground for realistic essentially 3 body studies of complex problems, aiding in identifying the most worthwhile problems.

Section~\ref{Sec2} develops a tractable formulation of the three-body Faddeev equation for the nucleon, including a description of the SCI and associated regularisation scheme.
Constraints imposed by S$_3$ permutation symmetry on the amplitude obtained as a solution of the Faddeev equation are described in Sect.\,\ref{S3FA}; and nucleon solutions of the SCI $3$-body equation are discussed in Sect.\,\ref{SecSol}.
The photon + nucleon interaction current appropriate for our approach to the nucleon bound state is described in Sect.\,\ref{NCurrent}.
Section~\ref{SecFFs} reports and discusses results obtained for nucleon electromagnetic form factors using that current.
A summary and perspective are provided in Sect.\,\ref{Epilogue}.

\section{SCI Faddeev Equation}
\label{Sec2}
\subsection{Tractable formulation}
In approaching any continuum bound state problem in quantum field theory, the first step is to decide upon the approximation that will be used to develop a tractable set of bound-state equations.
We choose to work at leading-order in a systematically-improvable, symmet\-ry-preserving truncation of all Dyson-Schwinger (quantum field) equations which are required for analysis of the $3$-body Faddeev equation and photon + nucleon interaction.
This is the rainbow-ladder (RL) truncation \cite{Munczek:1994zz, Bender:1996bb}.

After around thirty years of use, RL truncation is known to be reliable for pion, kaon, and nucleon observables:
(\emph{i}) practically, via numerous successful applications \cite{Roberts:2021nhw, Ding:2022ows, Raya:2024ejx};
and (\emph{ii}) because improvement schemes exist, with comparisons showing that the cumulative effect of amendments to RL truncation in these channels can be absorbed into a modest modification of the quark + quark scattering kernel \cite{Fischer:2009jm, Chang:2009zb, Chang:2013pq, Binosi:2014aea, Williams:2015cvx, Binosi:2016rxz, Binosi:2016wcx, Xu:2022kng}.
Where sensible comparisons are possible, modern CSM predictions obtained in this way are confirmed by  contemporary lattice-QCD results; see, \emph{e.g}., Refs.\,\cite{Roberts:2021nhw, Binosi:2022djx, Ding:2022ows, Ferreira:2023fva, Raya:2024ejx, Chen:2021guo, Chang:2021utv, Lu:2023yna, Chen:2023zhh, Yu:2024qsd, Alexandrou:2024zvn}.
 Consequently, comparisons between existing realistic $3$-body results and those obtained with the SCI reveal the sensitivity of a given body of observables to the momentum dependence of the quark-quark interaction.
This feature has already been exploited in connection with $q(qq)$ studies of nucleon properties; see, \emph{e.g}., Refs.\,\cite{Roberts:2011cf, Wilson:2011aa, Segovia:2013rca} \emph{cf}.\ Refs.\,\cite{Chen:2019fzn, Segovia:2015hra, Segovia:2013rca}.

\begin{figure}[t]
\centerline{%
\includegraphics[clip, width=0.44\textwidth]{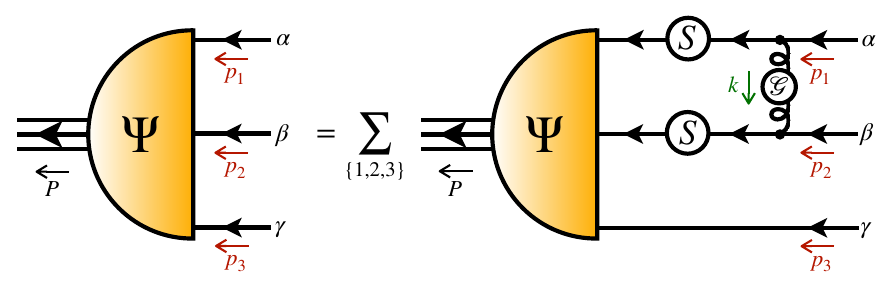}}
\caption{\label{FigFaddeev}
Rainbow-ladder truncation Faddeev equation.
Filled circle: Faddeev amplitude, $\Psi$, the matrix-valued solution.
Spring: dressed-gluon interaction that mediates quark+quark scattering; see Eqs.\,\eqref{EqRLInteraction}, \eqref{defcalG}.
Solid line: dressed-quark propagator, $S$, calculated from the rainbow gap equation.
Lines not carrying a shaded circle are amputated.
Isospin symmetry is assumed.
The sum runs over each of the cases involving quark ``$i=3,1,2$'' as a spectator to the exchange interaction.}
\end{figure}

Working in RL truncation, the nucleon bound state amplitude (amputated wave function) is obtained by solving the Faddeev equation sketched in Fig.\,\ref{FigFaddeev}.
Two inputs are required to complete the definition of the Faddeev equation:
(\emph{a}) ${\mathpzc G}$ -- the quark-quark interaction
and (\emph{b}) $S$ -- the dressed quark Schwinger function (propagator).

\subsection{SCI background}

Regarding (\emph{a}), in general, the RL interaction can be written as follows \cite{Maris:1997tm}:
{\allowdisplaybreaks
%%\begin{subequations}
%%\label{EqRLInteraction}
%%\begin{align}
%%\label{KDinteraction}
%%\mathscr{K}_{tu}^{rs}(k) & = {\mathpzc G}_{\mu\nu}(k) [i\gamma_\mu\frac{\lambda^{a}}{2} ]_{ts} [i\gamma_\nu\frac{\lambda^{a}}{2} ]_{ur}\,,\\
%
%% {\mathpzc G}_{\mu\nu}(k)  & = \tilde{\mathpzc G}(y) T_{\mu\nu}(k)\,,
%%\end{align}
%%\end{subequations}
\begin{equation}
\label{EqRLInteraction}
%% r s t u
%% A B C D
\mathpzc G_{CD}^{AB}(k)  =
\tilde{\mathpzc G}_{\mu\nu}(k)
 [i\gamma_\mu\frac{\lambda^{a}}{2} ]_{CB} [i\gamma_\nu\frac{\lambda^{a}}{2} ]_{DA}\,,
 %T_{\mu\nu}^k\,,
\end{equation}
where
% $k^2 T_{\mu\nu}^k = k^2 \delta_{\mu\nu} - k_\mu k_\nu$,  $y=k^2$, and
$A, B, C, D$ express colour and spinor matrix indices.
%The $T_{\mu\nu}$ specifies Landau gauge, used because it is a fixed point of the renormalisation group and also the gauge for which corrections to RL truncation are least significant \cite{Bashir:2009fv}.
%
}
A realistic form of $\tilde{\mathpzc G}_{\mu\nu}(y)$, constructed with reference to QCD's gauge sector, is discussed in Refs.\,\cite{Qin:2011dd, Binosi:2014aea}.
The SCI, on the other hand, is defined by writing \cite[Appendix A]{Cheng:2024gyv}:
\begin{equation}
\label{defcalG}
 \tilde{\mathpzc G}_{\mu\nu}  = \delta_{\mu\nu}\frac{4\pi a_{\rm{IR}}}{m_G^2}\,,
\end{equation}
where $m_G = 0.5\,$GeV is the gluon mass scale that is dynamically generated by gluon self-interactions \cite{Cui:2019dwv, Deur:2023dzc} and $\alpha_{\rm IR}=0.36\pi$ sets the quark + quark scattering strength.
This $\alpha_{\rm IR}$ value was determined fifteen years ago \cite{Roberts:2011wy}, in a unifying study of $\pi$- and $\rho$-meson properties, and has thereafter remained unchanged, providing the foundation for numerous, efficacious SCI analyses; see, \emph{e.g}.,
Refs.\,\cite{Chen:2012txa, Serna:2017nlr, Zhang:2020ecj, Xing:2022sor, Cheng:2024gyv, Sultan:2024hep, Xing:2025eip, Gutierrez-Guerrero:2019uwa, Chen:2024emt, Gutierrez-Guerrero:2026rsb, Wilson:2011aa, Segovia:2013rca, Xu:2015kta, Yin:2019bxe, Raya:2021pyr, Cheng:2022jxe, Yu:2025fer}.

The other element in Fig.\,\ref{FigFaddeev} is the dressed quark propagator.
Using Eq.\,\eqref{EqRLInteraction}, $S$ is obtained by solving the SCI rainbow-truncation gap equation \cite{Gutierrez-Guerrero:2010waf}:
\begin{align}
\label{GapEqn}
S^{-1}(k)  & = i\gamma\cdot k +m \nonumber \\
& \quad + \frac{16 \pi}{3} \frac{\alpha_{\rm IR}}{m_G^2}
\int \frac{d^4 l}{(2\pi)^4} \gamma_\mu S(l) \gamma_\mu\,,
\end{align}
where $m$ is the light-quark current-mass.
The integral on the right-hand side is divergent.  The regularisation scheme used to define such formulae is an essential part of the definition of the SCI.  It is detailed in \ref{AppendixA}.
Stated succinctly, divergent integrals are given meaning via a proper-time regularisation scheme \cite{Ebert:1996vx}, with an infrared cutoff that simulates confinement and an ultraviolet cutoff that sets the scale for all mass-dimensioned quantities and, implicitly, an upper bound on the domain of SCI validity.

The general form of the solution to Eq.\,\eqref{GapEqn} is
\begin{equation}
S(k) = Z/[i \gamma\cdot k + M]\,,
\end{equation}
where $Z$, $M$ are momentum-independent in the SCI.  Using any symmetry-preserving regularisation scheme, one finds by inspection that $Z=1$.  The dressed mass, on the other hand, is determined dynamically by solving
\begin{equation}
M = m +
\frac{16 }{3} \frac{4 \pi \alpha_{\rm IR}}{m_G^2}
\int \frac{d^4 l}{(2\pi)^4} \frac{M}{l^2+M^2} \,.
\label{gap1}
\end{equation}
Herein, to define the integral in Eq.\,\eqref{gap1}, we adopt the scheme described in \ref{AppendixA1}.  Using Eq.\,\eqref{CnIntegrals}, Eq.\,\eqref{gap1} can be written:
\begin{equation}
M = m + M \frac{4 \alpha_{\rm IR}}{3\pi m_G^2} {\mathsf C}_{2}^{\rm iu}(M^2)\,.
\end{equation}
Exploiting Eq.\,\eqref{OldRegNew}, this equation is seen to be identical to Ref.\,\cite[Eq.\,(A6)]{Cheng:2024gyv}: both schemes yield the same gap equation.
Similarly, they yield the same Bethe-Salpe\-ter equations.

\begin{table}[t]
\caption{ \label{SCIparamsResults}
SCI input parameters and some associated results.
As usual \cite[Appendix A]{Cheng:2024gyv}, $\alpha_{\rm IR} = 0.36\pi$, $m_G=0.5\,$GeV,
$\Lambda_{\rm ir} = 0.24\,$GeV ($1/\Lambda_{\rm ir} = 0.82\,$fm, which is a confinement scale in fair agreement with proton electromagnetic radii \cite{Cui:2022fyr}.)
$m$, $\Lambda_{\rm uv}$ are varied to obtain the listed values of $m_\pi$, $f_\pi$.
$\alpha_{\rm IR}^N$, dimensionless, is tuned to deliver the measured nucleon mass, $m_N=0.94\,$GeV.
%
%\textcolor[rgb]{1,0.00,0.00}{one loop 2 GeV masses}
%
(All dimensioned quantities listed in GeV.)
}
\begin{tabular*}
{\hsize}
{
l@{\extracolsep{0ptplus1fil}}
l@{\extracolsep{0ptplus1fil}}
l@{\extracolsep{0ptplus1fil}}
|l@{\extracolsep{0ptplus1fil}}
l@{\extracolsep{0ptplus1fil}}
l@{\extracolsep{0ptplus1fil}}
l@{\extracolsep{0ptplus1fil}}}\hline
\centering
%\begin{tabular}{lcccc}
$m\ $ & $\Lambda_{\rm uv}\ $ & $\alpha_{\rm IR}^N\ $ & $M\ $ & $m_\pi\ $ & $ f_\pi\ $ & $m_N\ $ \\\hline
$0.007\ $ & $0.91\ $ & $\alpha_{\rm IR}\ $ & $0.36\ $ & $0.14\ $ & $0.10\ $ & $0.67\ $\\
$0.007\ $ & $0.91\ $ & $\alpha_{\rm IR} \times 0.75 \ $ & $0.36\ $ & $0.14\ $ & $0.10\ $ & $0.94\ $\\\hline
\end{tabular*}
\end{table}

SCI input parameters have been fixed through a large body of comparisons with data; see., \emph{e.g}., Refs.\,\cite{Zhang:2020ecj, Xing:2022sor, Cheng:2024gyv, Sultan:2024hep, Xing:2025eip, Chen:2024emt, Wilson:2011aa, Segovia:2013rca, Xu:2015kta, Yin:2019bxe, Raya:2021pyr, Cheng:2022jxe, Yu:2025fer}.  The light-quark values are listed in Table~\ref{SCIparamsResults} along with some associated, derived results.

In closing this section, some remarks concerning the connection between SCI analyses and EHM may be useful.
(\emph{i}) The interaction in Eq.\,\eqref{defcalG} is infrared finite and characterised by a nonzero gluon mass scale.  These are the first two pillars of EHM \cite{Ding:2022ows}.
(\emph{ii}) The gap equation built from these inputs takes a light quark current mass, $m$, and dynamically transforms it into a dressed mass $M \approx 50 \times m$.  This magnification is the third pillar of EHM.
With these properties, the SCI expresses the same qualitative features as QCD.
The crucial differences are that the SCI gluon mass, quark-quark coupling, and dressed quark mass are momentum independent, whereas the QCD analogues are strongly momentum dependent at momenta $|k|> M$.
Consequently, as demonstrated in numerous applications -- see, \emph{e.g}., Refs.\,\cite{Chen:2012txa, Yin:2019bxe, Gutierrez-Guerrero:2019uwa, Xing:2022sor}, SCI predictions are phenomenologically reliable for long-wavelength observables, but must otherwise differ from those produced by QCD.
Thus, the comparisons we subsequently provide between SCI $3$-body results and those obtained with a QCD-connec\-ted interaction serve to highlight the impacts of QCD-running on observable hadron properties; hence, may prove useful in using experiment to map the true quark-quark interaction.

\subsection{Mathematical expression of SCI Faddeev equation}
\label{SubSecFE}
At this point, all inputs to the Faddeev equation sketched in Fig.\,\ref{FigFaddeev} are available.  Therefore, it is here appropriate to give mathematical meaning to that image.  Using the SCI, the solution is independent of quark + quark relative momenta:
\begin{align}
\label{faveq}
\Psi_{A B}^{C D}(P)&= \sum_{i=1}^3\Psi^{(i)}_{A B C D}(P)\,,
\end{align}
where $P$ is the nucleon total momentum, $P^2 = -m_N^2$,
{\allowdisplaybreaks
\begin{subequations}
\begin{align}
		\Psi^{(1)}_{A B C D}(P)&= \int_{\sf dk}\mathcal{G}^{C^\prime B^\prime}_{B\ C}(k) {S}_{B^{\prime} B^{\prime \prime}}(k_{2}) {S}_{C^{\prime} C^{\prime \prime}}(\tilde{k}_{3})
\nonumber \\
		&\quad \times \Psi_{A B^{\prime \prime}}^{C^{\prime \prime} D}(P)\,,\\
		\Psi^{(2)}_{A B C D}(P)& = \int_{\sf dk}\mathcal{G}^{A^\prime C^\prime}_{C\ A}(k){S}_{C^{\prime} C^{\prime \prime}}(k_{3}) {S}_{A^{\prime} A^{\prime \prime}}(\tilde{k}_{1}) \nonumber\\
		&\quad \times\Psi_{A^{\prime \prime} B}^{C^{\prime \prime} D}(P)\,, \\
		\Psi^{(3)}_{A B C D}(P) & =\int_{\sf dk}\mathcal{G}^{B^\prime A^\prime}_{A\ B}(k){S}_{A^{\prime} A^{\prime \prime}}(k_{1}) {S}_{B^{\prime} B^{\prime \prime}}(\tilde{k}_{2})\nonumber \\
& \quad \times\Psi_{A^{\prime \prime} B^{\prime \prime}}^{C D}(P)\,.
\label{faveq3c}
	\end{align}
	\label{faveq3}
\end{subequations}
\hspace*{-0.25\parindent}In Eq.\,\eqref{faveq3},
$\int_{\sf dk}$ represents the translationally invariant SCI regularisation of the four-dimensional momentum-space integral -- \ref{AppendixA1} -- and the meaning of the indices $A, B, C, D$ has been extended to include spinor ($\alpha, \beta, \gamma, \delta$), colour ($r, s, t, u$), and flavour ($a, b, c, d)$, with $D$ labelling these quantities for the nucleon itself.
}

The quark propagators in Eq.\,\eqref{faveq3} depend on the independent particle momenta $p_{1}$, $p_{2}$, $p_{3}$: momentum conservation requires $p_1+p_2+p_3=P$.
Following, \emph{e.g}., Ref.\,\cite{Xu:2025ups} and accommodating Eq.\,\eqref{faveq}, we write
$k_{1,2,3} = (1/3) P - k$, $\tilde k_{1,2,3} = (1/3) P + k$, where $k$ is the loop integration (exchange) momentum.
Here, the multiplicative constant $(1/3)$ reflects a choice of momentum sharing parameter.
Using any symmetry preserving regularisation scheme, results for observables are independent of this choice, but equal sharing is useful because it simplifies analyses.

Given the spin, flavour, and colour structure of the nucleon, the amplitude in Eqs.\,\eqref{faveq}, \eqref{faveq3} can be decomposed as follows:
\begin{align}
	\Psi_{A B}^{C D}(P)&=
\frac{\epsilon_{rst}}{\surd 6} \otimes
\sum_{\rho=0}^1 \psi_{\alpha\beta\gamma}^{\rho \, {\cal I}}(P) \otimes
{\mathrm F}^{\rho}_{abcd} \,,
	\label{amp}
\end{align}
where the colour factor $\epsilon_{rst}$ implements antisymmetry of the nucleon amplitude under interchange of any two dressed-quark constituents, so the overall nucleon colour label is $u=0$.

{\allowdisplaybreaks
The remaining elements possess S$_3$ permutation symmetry.  To be explicit, we represent the flavour amplitude in an isospin basis spanned by the
mixed–antisym\-metric (MA) and mixed–sym\-metric (MS) combinations:
\begin{subequations}
\label{FlavourM}
\begin{align}
\mbox{\rm MA:} \quad F^{0}_{abcd} &= \frac{1}{\surd{2}}\, i \left[\sigma_2\right]_{ab} \left[\mathbb{I}\right]_{cd} \,,\\
\mbox{\rm MS:} \quad F^{1}_{abcd} &= -\frac{1}{\surd{6}}\, \left[\sigma_j i\sigma_2 \right]_{ab} \left[\sigma_j \right]_{cd},
\end{align}
\end{subequations}
where $[\mathbb{I} = {\rm diag}[1,1]$ and $\{ \sigma_i, i=1,2,3\}$ are the Pauli matrices.
Under $a\leftrightarrow b$, the MA term in Eq.\,\eqref{FlavourM} is antisymmetric and MS is symmetric.
If trying to draw parallels with a $q(qq)$ picture \cite{Barabanov:2020jvn}, then MA corresponds to the isoscalar-scalar channel and MS to isovector-axial\-vector.
One may readily establish the following orthonormality relations:
\begin{equation}
\label{FON}
[F^{\rho^\prime \dagger}]_{bad^\prime c} [F^{\rho}]_{bacd}  =
\delta_{\rho^\prime \rho} \delta_{d^\prime d}\,.
\end{equation}
}

The remaining element in Eq.\,\eqref{amp} is $\psi_{\alpha\beta\gamma}^{\rho \, {\cal I}}$, the momen\-tum-spin component, with ${\cal I}$ recording the spin of the nucleon and $\rho = 0,1$ ranges over MA, MS.
Were the quark-quark interaction to be momentum dependent, then $64=32\times 2$ Dirac matrix valued tensors \cite[Table~2]{Eichmann:2009en} would be necessary to specify $\psi_{\alpha\beta\gamma}^{\rho \, {\cal I}}$ completely.
Using the SCI, on the other hand, only $8=4\times 2$ are necessary:
\begin{align}
\label{basise}
\psi^{\rho\, {\cal I}}_{\alpha \beta \gamma}(P)&:=
\sum_{i=1}^{4} \sum_{\ell=\pm} f^{\rho  i  \ell}(P)
{\mathsf X}_{\alpha \beta \gamma {\cal I}}^{i \ell} .
\end{align}

Here, the expansion coefficients, $f^{\rho  i  \ell}$, are determined by solving the SCI Faddeev equation: they are  independent of quark + quark relative momenta.
The basis vectors are:
\begin{subequations}
\label{Xbasis}
\begin{align}
{\mathsf X}_{\alpha \beta \gamma {\cal I}}^{1 \ell}
& = {\mathsf S}_{\alpha \beta \gamma {\cal I}}^{\ell}
= [\Lambda_\ell \gamma_5 C]_{\alpha \beta} [\Lambda_+]_{\gamma {\cal I}}\,, \\
{\mathsf X}_{\alpha \beta \gamma {\cal I}}^{2 \ell}
& = {\mathsf A}_{\alpha \beta \gamma {\cal I}}^{\ell}
=\tfrac{1}{\surd 3}  [\gamma_5 \gamma_\mu^T\Lambda_\ell \gamma_5 C]_{\alpha \beta} [\gamma_5 \gamma_\mu^T\Lambda_+]_{\gamma {\cal I}}\,, \\
{\mathsf X}_{\alpha \beta \gamma {\cal I}}^{3 \ell}
& = {\mathsf P}_{\alpha \beta \gamma {\cal I}}^{\ell}
= [\gamma_5  \Lambda_\ell \gamma_5 C]_{\alpha \beta} [\gamma_5\Lambda_+]_{\gamma {\cal I}}\,, \\
{\mathsf X}_{\alpha \beta \gamma {\cal I}}^{4 \ell}
& = {\mathsf V}_{\alpha \beta \gamma {\cal I}}^{\ell}
=\tfrac{1}{\surd 3}  [\gamma_\mu^T\Lambda_\ell \gamma_5 C]_{\alpha \beta} [ \gamma_\mu^T\Lambda_+]_{\gamma {\cal I}}\,,
\end{align}
\end{subequations}
where $\Lambda_\pm = [M \mp i \gamma\cdot P]/[2 M]$ are positive/negative energy projection operators,
$C = \gamma_2\gamma_4$ is the charge conjugation matrix,
and $\gamma_\mu^T = T_{\mu\nu}^P \gamma_\nu$, $T_{\mu\nu}^P  = \delta_{\mu\nu} - P_\mu P_\nu/P^2$.
In each case, the trailing $\Lambda_+$ projects the general Dirac structure onto that of a positive energy nucleon.

The basis vectors are orthonormalised:
\begin{align}
	\label{orth}
\frac{1}{4} {\rm tr} \,
[\overline{\mathsf X}_{\beta\alpha \delta \gamma}^{i \ell} {\mathsf X}_{\alpha \beta \gamma \delta}^{j \ell^\prime}] = \delta_{i j} \delta_{\ell \ell^\prime}\,.
\end{align}
Now, with reference to Eq.\,\eqref{Xbasis}, we write (no  sum on $i$)
\begin{align}
{\mathsf X}_{\alpha \beta \gamma \delta}^{i \ell}(P)
& = {\mathpzc I}_{\alpha \beta}^{i \ell}(P) {\mathpzc O}_{\gamma \delta}^i(P)\,,
\end{align}
in which case, one can express (with $[\cdot]^{\rm T}$ indicating matrix transpose)
\begin{align}
\bar {\mathsf X}_{\beta \alpha \delta \gamma }^{i \ell}&(P) \nonumber \\
& =
[ C^\dagger [{\mathpzc I}^{i \ell}(-P)]^{\rm T} C ]_{\beta\alpha }
[ C^\dagger [{\mathpzc O}^i(-P)]^{\rm T} C]_{\delta \gamma }\,.
\label{barX}
\end{align}

Now inserting Eqs.\,\eqref{EqRLInteraction},  \eqref{amp} into Eq.~\eqref{faveq3c}, one obtains
\begin{equation}\label{eq:psi3-spinmom}
\begin{aligned}
\psi^{\rho\,{\mathcal I}\,(3)}_{\alpha\beta\gamma}(P)
&= \frac{2}{3}
\int_{\sf dk} \frac{4\pi \alpha_{\rm IR}}{m_G^2}
[\gamma_\mu]_{\alpha\alpha'}S_{\alpha'\alpha''}(k_1)[\gamma_\mu]_{\beta\beta'}\\
&S_{\beta'\beta''}(\tilde k_2)
\psi^{\rho\,\mathcal I}_{\alpha''\beta''\gamma}(P).
\end{aligned}
\end{equation}
Here, the colour and flavour projections have explicitly been completed using Eq.\,\eqref{FON}.

For many purposes in concrete calculations, it is useful to take advantage of the S$_3$ permutation symmetry of the complete Faddeev amplitude.
To that end, for clarity, we first explicitly express the six permutations of the flavour-spin structure element in the Eq.\,\eqref{amp} amplitude:
\begin{subequations}
\begin{align}
\mbox{$(123)$:} & \; [F^\rho] = F^{\rho}_{abcd}\,,\; [\psi^\rho]= \psi^{{\cal I}\rho}_{\alpha\beta\gamma}\,;\\
\mbox{$(231)$:} & \; [F^{\rho\prime}] = F^{\rho}_{bcad}\,,\; [\psi^{\rho\prime}]= \psi^{{\cal I } \rho}_{\beta\gamma\alpha}\,;\\
\mbox{$(312)$:} & \; [F^{\rho\prime\prime}] = F^{\rho}_{cabd}\,,\; [\psi^{\rho\prime\prime}]= \psi^{{\cal I } \rho}_{\gamma\alpha\beta}\,;\\
\mbox{$(213)$:} & \; [\tilde{F}^{\rho}] = F^{\rho}_{bacd}\,,\; [\tilde\psi^{\rho}]= \psi^{{\cal I } \rho}_{\beta\alpha\gamma}\,;\\
\mbox{$(321)$:} & \; [\tilde F^{\rho\prime}] = F^{\rho}_{cbad}\,,\; [\tilde\psi^{\rho\prime}]= \psi^{{\cal I } \rho}_{\gamma\beta\alpha}\,;\\
\mbox{$(132)$:} & \; [\tilde F^{\rho\prime\prime}] = F^{\rho}_{acbd}\,,\; [\tilde\psi^{\rho\prime\prime}]= \psi^{{\cal I } \rho}_{\alpha\gamma\beta}\,,
\end{align}
\end{subequations}

It is also convenient to employ the following compact notation for the flavour-spin structure in Eq.\,\eqref{amp}:
\begin{subequations}
\begin{align}
\psi^0 F^0 + \psi^1 F^1 & =
\sum_{a=3,1,2} [\psi^{0(a)} F^0 + \psi^{1(a)} F^1 ]\,.
\end{align}
This can be written in column vector form:
\begin{align}
\left(
\begin{array}{c}
  \psi^0 \\ \psi^1
\end{array}\right)
& = \sum_{a=3,1,2}
\left(\begin{array}{c}
  \psi^{0(a)} \\ \psi^{1(a)}
\end{array}\right)\\
& =
\left(
\begin{array}{c}
  \psi^{0(3)} \\ \psi^{1(3)}
\end{array}\right)
+
\left(
\begin{array}{c}
  \psi^{0(1)} \\ \psi^{1(1)}
\end{array}\right)
+
\left(
\begin{array}{c}
  \psi^{0(2)} \\ \psi^{1(2)}
\end{array}\right)\,.
\end{align}
\end{subequations}

S$_3$ invariance of the amplitude $\Psi$ in Eq.\,\eqref{amp} can now be seen to entail:
\begin{subequations}
\begin{align}
\psi_{\alpha\beta\gamma}^{\cal I}
& = \left(
\begin{array}{c}
  \psi_{\alpha\beta\gamma}^{0{\cal I}} \\
  \psi_{\alpha\beta\gamma}^{1{\cal I}}
\end{array}
\right)
 = {\mathpzc M}^\prime \psi_{\beta\gamma\alpha}^{\cal I\prime }
= {\mathpzc M}^{\prime\prime} \psi_{\gamma\alpha\beta}^{\cal I\prime\prime } \,, \\
&
= \tilde{\mathpzc M} \tilde\psi_{\beta\alpha\gamma}^{\cal I}
= \tilde {\mathpzc M}^{\prime} \tilde\psi_{\gamma\beta\alpha}^{\cal I\prime }
= \tilde {\mathpzc M}^{\prime\prime} \tilde \psi_{\alpha\gamma\beta}^{\cal I\prime\prime } \,, \label{Eq18b}
\end{align}
\end{subequations}
where
\begin{align}
{\mathpzc M}^\prime & =
\tfrac{1}{2}\left(
\begin{array}{cc}
-1 & -\surd 3 \\
\surd 3 & -1
\end{array}
\right)\,, \quad
{\mathpzc M}^{\prime\prime} =
\tfrac{1}{2}\left(
\begin{array}{cc}
-1 & \surd 3 \\
-\surd 3 & -1
\end{array}
\right)\,,
\label{matricesM}
\end{align}
$\tilde{\mathpzc M}={\rm diag}[-1,1] $,
$\tilde {\mathpzc M}^\prime = \tilde {\mathpzc M} [{\mathpzc M}^\prime]^{\rm T}$,
$\tilde {\mathpzc M}^{\prime\prime} = \tilde {\mathpzc M} [{\mathpzc M}^{\prime\prime}]^{\rm T}$.
%with $[\cdot]^{\rm T}$ indicating matrix transpose.

%Consider now the flavour-spin product in Eq.\,\eqref{S$_3$ symmetry of the amplitude entails

Returning to Eq.\,\eqref{eq:psi3-spinmom}, one obtains the complete solution of the $3$-body Faddeev equation by inserting the following amplitude on the right-hand side:
\begin{equation}
\psi =
\left(
\begin{array}{c}
\psi^0 \\ \psi^1
\end{array}\right)
= \psi^{(3)} + {\mathpzc M}^\prime \psi^{(3)\prime} + {\mathpzc M}^{\prime\prime} \psi^{(3)\prime\prime}
\end{equation}
\emph{viz}.\ Eq.\,\eqref{eq:psi3-spinmom} is a closed equation for $\psi^{(3)}$; hence, by symmetry, for the complete amplitude.

The structure of the right-hand side of Eq.\,\eqref{eq:psi3-spinmom} can be represented as
\begin{equation}
{\mathpzc R}^{(3)} = \int_{\sf dk} \frac{{\mathpzc N}(k,P)}{[k_3^2+M^2][\tilde k_3^2+M^2]}\,,\label{EqFE1}
\end{equation}
where ${\mathpzc N}(k,P)$ is some numerator function.
Combining the denominators by using a Feynman para\-met\-risation, with variable ${\mathpzc a}$, and performing the momentum shift
$k \to k_{\mathpzc a} = k + (P/3)(2 {\mathpzc a}-1)$, Eq.\,\eqref{EqFE1} is recast into the form
\begin{equation}
{\mathpzc R}^{(3)} = \int_{\sf dk} \frac{{\mathpzc N}(k_{\mathpzc a},P)}{[k^2 + \sigma]^2 }\,,\label{EqFE2}
\end{equation}
where $\sigma = M^2 + {\mathpzc a}(1-\mathpzc a)(4/9)P^2 = M^2-{\mathpzc a}(1-\mathpzc a)(4/9)m_N^2$.

By inspection of Eq.\,\eqref{eq:psi3-spinmom}
%, performing a dimensional analysis,
and recognising that the denominator in Eq.\,\eqref{EqFE2} is an even function of the integration variable, $k$, one sees that the numerator can only involve $M^2$, $k^2$, $(k\cdot P)^2$, and products of these factors.
All such integrals can be rendered finite using the regularisation scheme described in \ref{AppendixA1}.
One therewith arrives at a set of $16$ coupled equations for the expansion coefficients in Eq.\,\eqref{basise}: $\{f^{\rho  i \ell}|\rho=1,2; i=1,2,3,4;\ell = \pm\}$.

\section{S$_3$ Symmetry and the Faddeev Amplitude}
\label{S3FA}
In connection with the expansion coefficients in Eq.\,\eqref{basise}, it is worth elucidating important consequences of S$_3$ symmetry.  Recall the following entry in Eq.\,\eqref{Eq18b}:
\begin{align}
  \psi_{\alpha\beta\gamma}^{\cal I} & =  \tilde{\mathpzc M} \tilde\psi_{\beta\alpha\gamma}^{\cal I}
  \leftrightarrow
  \left(\begin{array}{c}
  \psi_{\alpha\beta\gamma}^{0 \cal I}\\ \psi_{\alpha\beta\gamma}^{1 \cal I}
  \end{array}\right)
=   \left(\begin{array}{c}
  -\psi_{\beta\alpha\gamma}^{0 \cal I}\\ \psi_{\beta\alpha\gamma}^{1 \cal I}
  \end{array}\right)\,.
  \label{Exchange}
\end{align}
Expanding all elements via Eq.\,\eqref{basise};
multiplying on the left by
$\overline{\mathsf X}_{\beta\alpha {\cal I} \gamma}^{i^\prime \ell^\prime}$;
and using orthonormality, Eq.\,\eqref{orth}, one finds that Eq.\,\eqref{Exchange} (exchange symmetry) entails:
\begin{subequations}
\begin{align}
f^{02\pm} & = 0\,,\quad
f^{11\pm} = 0\,, \\
f^{03\pm} & = f^{03\mp}\,,\quad
f^{04\pm} = f^{04\mp}\,,\\
f^{13\pm} & =- f^{13\mp}\,, \quad
f^{14\pm} =  - f^{14\mp}\,.
\end{align}
\end{subequations}
At most, therefore, the SCI can support $8=(4-1)\times 2\times 2-4$ independent nonzero expansion coefficients.

Now consider the constraints imposed by cyclic symmetry:
\begin{align}
  \psi_{\alpha\beta\gamma}^{\cal I} & =  {\mathpzc M}^\prime \psi_{\beta\gamma\alpha}^{\cal I\prime } \nonumber \\
&  \leftrightarrow
  \left(\begin{array}{c}
  \psi_{\alpha\beta\gamma}^{0 \cal I}\\ \psi_{\alpha\beta\gamma}^{1 \cal I}
  \end{array}\right)
=   \left(\begin{array}{cc}
  -\tfrac{1}{2} \psi_{\beta\gamma\alpha}^{0 \cal I}
      & -\tfrac{\surd 3}{2} \psi_{\beta\gamma\alpha}^{1 \cal I} \\
  \tfrac{\surd 3}{2} \psi_{\beta\gamma\alpha}^{0 \cal I}
      & -\tfrac{1}{2} \psi_{\beta\gamma\alpha}^{1 \cal I} \\
  \end{array}\right)\,.
  \label{Cyclic}
\end{align}
Working as before and using Eq.\,\eqref{Exchange}, one obtains the following identities:
\begin{subequations}
\label{EqCyclic}
\begin{align}
f^{01-} & = f^{03+} + \surd 3 f^{04+} ,\; \quad
f^{01+} = f^{12+}\,,\\
%% f^{011} & = f^{032} + \surd 3 f^{042}
%
f^{03-}& = f^{03+} \,,\quad f^{04-} = f^{04+} \\
f^{12-} & = -  f^{03+}+\tfrac{1}{\surd 3} f^{04+} ,\\
f^{13\pm} & = \pm f^{04\pm} \,,\quad
f^{14\pm} = \pm f^{03\pm} \pm \tfrac{2}{\surd 3} f^{04\pm}.
\end{align}
\end{subequations}
%% I get f^{011}-sqrt[3]f^{111}=f^{011}+sqrt[3]f^{021}

\begin{table}[t]
\caption{ \label{nucleonresults}
SCI results for the nucleon mass and Faddeev amplitude, normalised for illustration by requiring
$\sum_{\rho=0,1}\sum_{i=1}^{4} \sum_{\ell = \pm} [f^{\rho i \ell}]^2 = 1$.
{\sf Panel A}. Quark + quark coupling in the Faddeev equation is the same as that in the gap and Bethe-Salpeter equations: $m_N = 0.67\,$GeV.
{\sf Panel B}. Reduced-strength quark + quark coupling in the Faddeev equation: $\alpha_{\rm IR}^N = 0.75\alpha_{\rm IR}$: $m_N=0.94\,$GeV.
%
%\textcolor[rgb]{1,0.00,0.00}{one loop 2 GeV masses}
%
%(All dimensioned quantities listed in GeV.)
}
\begin{tabular*}
{\hsize}
{
l@{\extracolsep{0ptplus1fil}}
|l@{\extracolsep{0ptplus1fil}}
l@{\extracolsep{0ptplus1fil}}
l@{\extracolsep{0ptplus1fil}}
l@{\extracolsep{0ptplus1fil}}
l@{\extracolsep{0ptplus1fil}}
l@{\extracolsep{0ptplus1fil}}
l@{\extracolsep{0ptplus1fil}}
l@{\extracolsep{0ptplus1fil}}}\hline
\centering
%\begin{tabular}{lcccc}
$\alpha_{\rm IR}\ $ & $f^{\rho1+}\ $ & $f^{\rho1-}\ $ & $f^{\rho2+}\ $ & $f^{\rho2-}\ $ & $f^{\rho3+}\ $ & $f^{\rho3-}\ $ & $f^{\rho4+}\ $ &$f^{\rho4-}\ $ \\\hline
$\rho = 0\ $ & $0.72\ $ & $0.53\ $ & $0\ $ & $0\ $ & $0\ $ & $\phantom{-}0\ $ & $0.31\ $ & $\phantom{-}0.31\ $\\
$\rho = 1\ $ & $0\ $ & $0\ $ & $0.72\ $ & $0.18\ $ & $0.31\ $ & $-0.31\ $& $0.36\ $ & $-0.36\ $ \\\hline
\end{tabular*}

\medskip

\begin{tabular*}
{\hsize}
{
l@{\extracolsep{0ptplus1fil}}
|l@{\extracolsep{0ptplus1fil}}
l@{\extracolsep{0ptplus1fil}}
l@{\extracolsep{0ptplus1fil}}
l@{\extracolsep{0ptplus1fil}}
l@{\extracolsep{0ptplus1fil}}
l@{\extracolsep{0ptplus1fil}}
l@{\extracolsep{0ptplus1fil}}
l@{\extracolsep{0ptplus1fil}}}\hline
\centering
%\begin{tabular}{lcccc}
$\alpha_{\rm IR}^N\ $ & $f^{\rho1+}\ $ & $f^{\rho1-}\ $ & $f^{\rho2+}\ $ & $f^{\rho2-}\ $ & $f^{\rho3+}\ $ & $f^{\rho3-}\ $ & $f^{\rho4+}\ $ &$f^{\rho4-}\ $ \\\hline
$\rho = 0\ $ & $0.70\ $ & $0.55\ $ & $0\ $ & $0\ $ & $0\ $ & $\phantom{-}0\ $ & $0.32\ $ & $\phantom{-}0.32\ $\\
$\rho = 1\ $ & $0\ $ & $0\ $ & $0.70\ $ & $0.18\ $ & $0.32\ $ & $-0.32\ $& $0.37\ $ & $-0.37\ $ \\\hline
\end{tabular*}
\end{table}

So, finally, accounting fully for S$_3$ symmetry, the SCI Faddeev equation supports just $3$ independent expansion coefficients, which may be chosen as:
\begin{equation}
f^{01+}\,,\; f^{03+}\,,\; f^{04+}\,.
\end{equation}
In physical terms, this result means that the SCI $3$-body nucleon Faddeev amplitude does not contain any term that depends on relative momentum, the positive and negative energy comments are intimately linked, and the MS flavour components are fully determined by the MA terms.

It is worth noting that any realistic-interaction $3$-body Faddeev equation also expresses identities such as these and more, too, because there are more Dirac matrix valued tensors in Eq.\,\eqref{basise}.

\section{Solution of SCI Faddeev Equation}
\label{SecSol}
Using the one- and two-body input parameters and associated derived quantities in Table~\ref{SCIparamsResults}, it is straightforward to solve the nucleon $3$-body Faddeev equation in Sect.\,\ref{SubSecFE}.  The results are listed in Table~\ref{nucleonresults}.

One first observes that if the SCI coupling used in the Faddeev equation is the same as that in the gap and Bethe-Salpeter equations, then the nucleon is overbound, \emph{viz}.\ the computed mass is just 71\% of the empirical value.  The associated amplitude is specified by the expansion coefficients in the upper panel of Table~\ref{nucleonresults}.

\begin{figure}[t]
\centerline{%
\includegraphics[clip, width=0.44\textwidth]{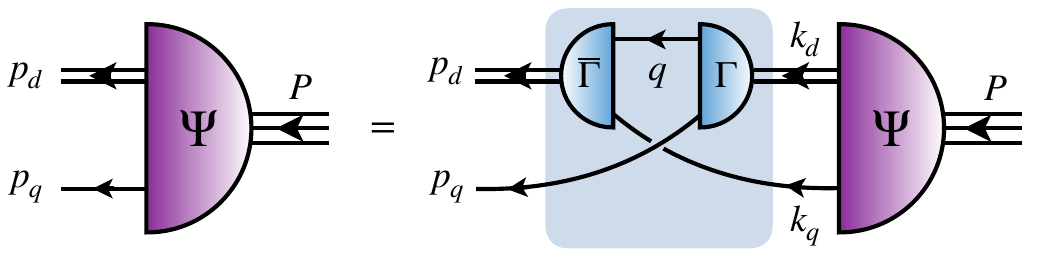}}
\caption{\label{qqFE}
Sketch of the $q(qq)$ Faddeev equation used, \emph{e.g}., in Refs.\,\cite{Roberts:2011cf, Xu:2015kta}.
$\Psi$ is the Poincar\'e-covariant solution amplitude for a baryon with total momentum $P=p_q+p_d=k_q+k_d$, constituted from three valence quarks, two of which are paired in a fully-interacting nonpointlike diquark correlation.
Here, $\Psi$ expresses the relative momentum correlation between the dressed-quarks and -diquarks.
Legend.
\emph{Single line} -- dressed-quark propagator;
$\Gamma$ -- diquark correlation amplitude;
\emph{double line} -- $(qq)$ (diquark) propagator;
and \emph{shaded box} -- Faddeev kernel, which explicitly shows the quark exchange binding mechanism, which is the horizonal quark line at the top of the diagram.
}
\end{figure}

In contrast, standard SCI treatments of the $q(qq)$ Faddeev equation -- see Fig.\,\ref{qqFE}, produce underbinding, \emph{i.e}., a nucleon mass in excess of the empirical value \cite{Roberts:2011cf, Xu:2015kta} unless the strength of the quark exchange kernel is magnified.

Given these observations, we accept some flexibility in the Faddeev equation ($3$-body) coupling.
Making the replacement $\alpha_{\rm IR} \to \alpha_{\rm IR}^N = 0.75 \alpha_{\rm IR}$, reflecting, in part, at least, the absence of spin-orbit repulsion in the SCI $3$-body system, one obtains $m_N=0.94$ and the amplitude in the lower panel of Table~\ref{nucleonresults}.
(This expedient is unnecessary when using a realistic interaction in the Faddeev equation; see, \emph{e.g}., Refs.\,\cite{Yao:2024uej, Yao:2024ixu}.)
Referring to Table~\ref{nucleonresults}, it is evident that the Faddeev amplitude is not particularly sensitive to the value of $\alpha_{\rm IR}$, with any reasonable value delivering a similar result.

%% These remarks already address the issues of physical interpretation (spin-orbit repulsion)
%   and predictive power (limited effect because the Faddeev amplitude is practically unchanged).
It is, perhaps, worth gathering these remarks into a succinct statement about the interpretation of the coupling rescaling and its impact on SCI predictions for the nucleon.
This is the first SCI analysis of the essentially $3$-body system, and one might have desired that the $2$-body coupling strength in the $3$ quark system would be the same as that in the mesonic $2$-body quark + antiquark channel.
However, a contact interaction suppresses orbital angular momentum in bound-state wave functions and there are many more nonzero orbital angular momentum, $L\neq 0$, components in a $3$-body system as compared to a $2$-body system.
So, a suppression of spin-orbit repulsion must have a larger impact in $3$-body systems.
Indeed, considering rest-frame orbital angular momentum in the realistic case, the nucleon has $112/2=56$-times more $L\neq 0$ components than the pion \cite{Eichmann:2009zx, Bhagwat:2006xi}.
Thus, it may be viewed as a good outcome that one can compensate for the relative underestimate of spin-orbit repulsion in the SCI nucleon be merely introducing a 25\% suppression of the $2$-body coupling.
As this expedient has practically no effect on the nucleon Faddeev amplitude, then its impact on SCI predictions is negligible.

\begin{figure*}[t]
\centerline{%
\includegraphics[clip, width=0.86\textwidth]{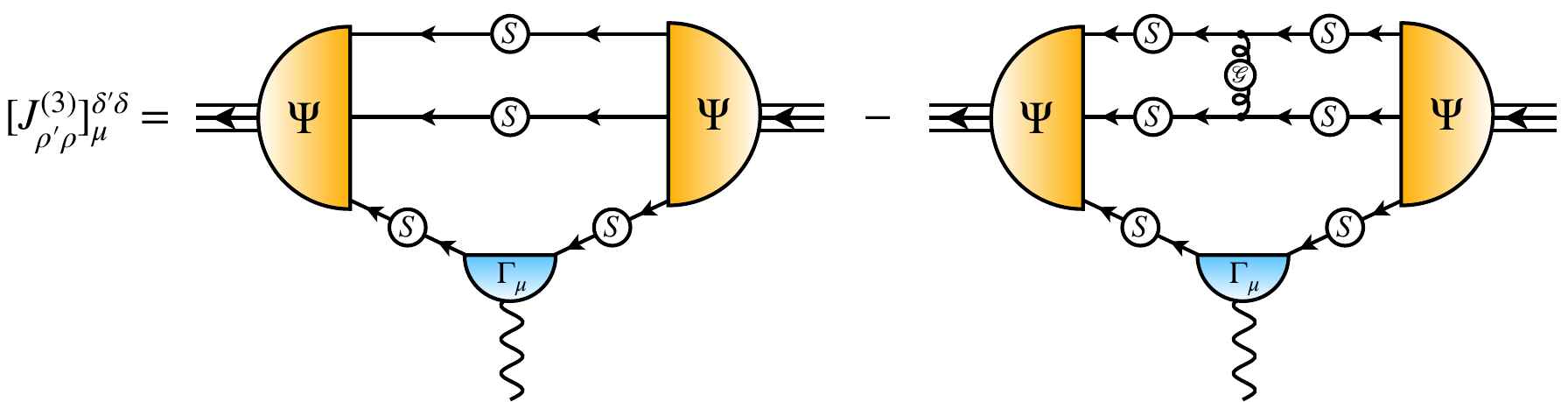}}
\caption{\label{FigCurrent}
The nucleon has three valence quarks; hence, the complete RL nucleon electromagnetic current has three terms: $J_\mu(Q) = \sum_{a=1,2,3} J_\mu^a(Q)$.
Symmetries mean that one can readily obtain the $a=1,2$ components once the $a=3$ component is known \cite[Appendix~B]{Eichmann:2011pv}.
%
%$a=3$ spinor component of the nucleon electromagnetic current.
$\delta$, $\delta^\prime$ are spinor indices and $\rho$, $\rho^\prime$ are isospin indices.
$\Gamma_\mu$ is the dressed-photon+quark vertex, Eq.\,\eqref{GammaQ} and, \emph{e.g}., Refs.\,\cite{Roberts:2011wy, Xu:2019ilh}.
}
\end{figure*}

Returning again to Table~\ref{nucleonresults}, one sees that, in all cases, the amplitude coefficients satisfy the identities derived in Sect.\,\ref{S3FA}.
The outcome $f^{03\pm}\equiv 0$ is peculiar to the SCI.  (One does not find this in realistic-interaction Faddeev equation solutions: therein, these amplitudes are small but nonzero.)
In this case, the identities in Sect.\,\ref{S3FA} entail: $f^{14+} = 2 f^{12-}$, which is satisfied.  Such comparisons are a useful check on the numerical solutions.

It is furthermore evident in Table~\ref{nucleonresults} that ${\mathsf X}^{1 \ell} = {\mathsf S}^{\ell}$, \emph{i.e}., scalar diquark like correlations, are dominant in the MA channel and ${\mathsf X}^{2 \ell} = {\mathsf A}^{\ell}$ (axialvector diquark like) correlations dominate in the MS channel.
Indeed, the $\Lambda_+$ components in both these channels are constrained by symmetry to have the same strength.  (This is also true in QCD-connected $3$-body Faddeev equation solutions \cite{Yao:2024uej}.)
In such outcomes, the SCI, too, delivers strong arguments against scalar-diquark-only models of proton structure.
Plainly, axialvector diquark like correlations are an essential feature of the proton wave function: this is an unavoidable consequence of Poincar\'e covariance.

\section{Nucleon Form Factors: Formulae}
\label{NCurrent}
%\subsection{Electromagnetic currents}
The photon interaction current for a nucleon whose structure is described by the solution of the Faddeev equation, Fig.\,\ref{FigFaddeev}, is obtained from the Schwinger function depicted in Fig.\,\ref{FigCurrent}.
Its general form is:
{\allowdisplaybreaks
\begin{align}
J_\mu^N(Q) & = ie \Lambda_+(P_f)
[ F_1^N(Q^2) \gamma_\mu \nonumber \\
& \quad + \frac{1}{2 m_N} \sigma_{\mu\nu} Q_\nu F_2^N(Q^2) ]
\Lambda_+(P_i) \,,
\label{NucleonCurrent}
\end{align}
where
the incoming and outgoing nucleon momenta are $P_{i,f}$, $P=(P_f+P_i)/2$, $Q=P_f-P_i$, $P_{i,f}^2=-m_N^2$,
$\Lambda_+(P_{i,f})$ are positive-energy nucleon-spinor projection operators,
$e$ is the positron charge,
and $F_{1,2}^N$ are the Dirac and Pauli form factors.
}

As usual \cite{Sachs:1962zzc}, the nucleon charge and magnetisation distributions are defined as  ($\tau = Q^2/[4 m_N^2]$):
\begin{equation}
G_E^N  = F_1^N - \tau F_2^N\,,
\quad G_M^N  = F_1^N + F_2^N \,.
\label{Sachs}
\end{equation}
Magnetic moments and radii are calculated therefrom:
$\mu_N = G_M^N(Q^2=0)$\,;
\begin{align}
\label{StandardDef}
\langle r_F^2\rangle^N & = \left. -6 \frac{d \ln G_F^N(Q^2)}{dQ^2}\right|_{Q^2=0}\,,
\end{align}
$F=E$, $M$, except $\langle r_E^2\rangle^n = -6 G_E^{n\prime}(Q^2)|_{Q^2=0}$ because $G_E^{n}(0)=0$.

In utilising the nucleon current in Fig.\,\ref{FigCurrent}, we follow the computational procedure described in Ref.\,\cite[Appendix~B]{Eichmann:2011pv}.
The electromagnetic current is expressed in the form
\begin{equation}
J_{\mu}^N(Q):=
\sum_{a=1}^{3}
J_\mu^{aN}(Q) =
\sum_{a=1}^3\sum_{\rho^{\prime} \rho}\mathsf{F}_{\rho^{\prime} \rho}^{(a)N}[J_{\rho^{\prime} \rho}^{(a) N)}]_{\mu}\,.
\label{eq:Jcurrent}
\end{equation}
Owing to permutation symmetries, it is sufficient to consider
\begin{equation}
\label{flavor_defF}
\mathsf{F}^{(3)N}_{\rho^\prime \rho}
=(e_N)_{d'}\,
(F^{\rho^\prime})^\dagger_{bad^\prime c^\prime }\,
{\cal Q}_{c'c}\,
F^\rho_{abcd}\,
(e_N)_d,
\end{equation}
where ${\cal Q}=\mathrm{diag}[q_u=2/3,q_d=-1/3]$ is the quark charge matrix,
and $e_{N=p}=(1,0)$, $e_{N=n}=(0,1)$ project, respectively, onto the proton and neutron states.
It is apparent from Eq.\,\eqref{flavor_defF} that the four elements in $\{\mathsf{F}^{(3)N}_{\rho^\prime \rho}\}$, labelled by $\{\rho^\prime, \rho\}$, are $2\times 2$ matrices in isospin space.

Physical proton and neutron currents (so, their form factors) are obtained by combining the $\rho=0,1$ contributions via the matrices \(\mathsf{F}_{\rho^\prime \rho}^{(a)}\) in Eq.\,\eqref{eq:Jcurrent}.
Evaluating the expressions in Eq.\,\eqref{flavor_defF}, one has:
\begin{equation}\label{flavor_p}
\mathsf{F}^{(3)p}  =
\begin{pmatrix}
  \tfrac{2}{3} & 0\\
  0 & 0
\end{pmatrix},
\qquad
\mathsf{F}^{(3)n} =
\begin{pmatrix}
 -\tfrac{1}{3} & 0\\
  0 & \tfrac{1}{3}
\end{pmatrix}.
\end{equation}

It remains only to write the mathematical expression for the photon + quark coupling in Fig.\,\ref{FigCurrent}:
\begin{align}
\label{ImFF}
 [J_{\rho^{\prime} \rho}^{(3)}]_{\mu}^{\delta^{\prime} \delta}&
  =\int_{\sf dp \sf dq}\bar{\psi}^{\rho^\prime \delta^{\prime} }_{\beta^{\prime} \alpha^{\prime} \gamma^{\prime}} \left(P_f\right)
 S_{\alpha^{\prime} \alpha}\left(p_1\right) S_{\beta^{\prime} \beta}\left(p_2\right)\nonumber \\
&  \times \chi_{\gamma^{\prime} \gamma}(p_3,Q)  \left[\psi^{\rho \delta }_{\alpha \beta \gamma}(P_i)
- \psi^{\rho \delta (3)}_{\alpha \beta \gamma}(P_i)\right],
\end{align}
where, working in the Breit frame, the internal quark momenta are
\begin{subequations}
\label{p1p2p3}
\begin{align}
p_1 & =-q - p/2 +P/3 \,, \quad p_2 = q - p/2 +P/3\,, \\
p_3^\pm & = p_3 \pm Q/2\,, \quad p_3= p+P/3 \,.
\end{align}
\end{subequations}
Regarding the conjugate amplitude, we have, see\linebreak Eq.\,\eqref{barX}:
\begin{align}
\label{basisebar}
\bar{\psi}^{\rho^\prime \delta^{\prime} }_{\beta^{\prime} \alpha^{\prime} \gamma^\prime} (P)
&=
\sum_{i=1}^{4} \sum_{\ell=\pm} f^{\rho^\prime, i, \ell}(P)
\bar {\mathsf X}_{\beta^{\prime} \alpha^{\prime}\delta^\prime\gamma^\prime}^{i\ell}(P)\,.
\end{align}

%%\begin{align}
%%\label{basise}
%%\psi^{\rho\, {\cal I}}_{\alpha \beta \gamma}(P)&:=
%%\sum_{i=1}^{4} \sum_{\ell=\pm} f^{\rho, i, \ell}(P)
%%{\mathsf X}_{\alpha \beta \gamma {\cal I}}^{i \ell} .
%%\end{align}

%$\chi(p_3,Q) = S(p_3^+)\Gamma_{\mu}(Q) S(p_3^-)$

The hitherto unspecified element in Eq.\,\eqref{ImFF} is\linebreak $\chi(p_3,Q) = S(p_3^+)\Gamma_{\mu}(Q) S(p_3^-)$, where $\Gamma_\mu(Q)$ is the dres\-sed photon + quark vertex.  This vertex was analysed in Ref.\,\cite{Roberts:2011wy}, with the result
\begin{subequations}
\label{GammaQ}
\begin{align}
\Gamma_\mu(Q) & = P_T(Q^2) \gamma_\mu^T + \gamma_\mu^L
=: \Gamma_\mu^T(Q) + \Gamma_\mu^L(Q) \,, \\
P_T(Q^2) & = \frac{1}{1+K_\gamma(Q^2)} \,, \\
K_{\gamma}(Q^2) & =
\frac{4}{3\pi} \frac{\alpha_{\rm{IR}}}{m_G^2}
\int_{0}^{1}d{\mathpzc a}\, {\mathpzc a}(1-{\mathpzc a})
Q^2 {\mathsf C}_0(\omega)\,,
\end{align}
\end{subequations}
where $\gamma_\mu^T = T_{\mu\nu}^Q\gamma_\nu$,
$\gamma_\mu^L = \gamma_\mu - \gamma_\mu^T$,
$\omega = M^2+ {\mathpzc a}(1-{\mathpzc a}) Q^2$.
A Ward-Green-Takahashi identity entails that only $\Gamma_\mu^T(Q)$ contributes to the form factors calculated herein.
Owing to the presence of $K_{\gamma}(Q^2)$, this vertex contains the $\rho$-meson pole that is commonly connected with vector meson dominance \cite{Xu:2021mju}.
However, in being obtained using RL truncation, it lacks the EHM-induced anomalous magnetic moment term \cite{Singh:1985sg, Bicudo:1998qb, Chang:2010hb, Qin:2013mta} that can make a noticeable contribution to bound-state magnetic moments \cite{Segovia:2014aza}.  The impact of this RL-truncation failing (not SCI) will become apparent below.

Using Eqs.\,\eqref{eq:Jcurrent}, \eqref{flavor_p} and S$_3$ permutation symmetry, one finds:
\begin{subequations}
\label{CurrentNucleon}
\begin{align}
\mathrm{proton:}\quad  & J^{\delta^{\prime} \delta ; p }_{\mu} =[2 J_{00}^{(3)}]^{\delta^{\prime} \delta}_{\mu},\\
\mathrm{neutron:} \quad  &J^{\delta^{\prime} \delta; n}_{\mu} =[J_{11}^{(3)}-J_{00}^{(3)}]^{\delta^{\prime} \delta}_{\mu}.
\end{align}
\end{subequations}
Reviewing Eqs.\,\eqref{FlavourM}, \eqref{flavor_defF}, it becomes clear that Eq.\,\eqref{CurrentNucleon} entails that there is no contribution to charged spin-half baryon form factors from MS terms in the bound-state amplitude.
This outcome is a feature of all such $3$-body treatments.  It is not specific to the SCI.
A similar result is expressed in $q(qq)$ pictures of nucleon structure \cite[Appendix\,C.1]{Wilson:2011aa}:
the electromagnetic form factors of such baryons do not receive any contributions associated with ``elastic'' scattering from the axialvector diquark piece of their Faddeev wave functions.

Working from Eqs.\,\eqref{CurrentNucleon}, one may compute the nucleon Sachs form factors (the trace is over Dirac indices):
\begin{subequations}
\label{eq:GEGM}
\begin{align}
G_E^N(Q^2) &=
\frac{1}{2i\sqrt{1+\tau}}\,
{\rm tr} \left[J^{N}_\mu \hat{P}_{\mu}\right], \\
G_M^N(Q^2) &= \frac{i}{4\tau}\,
{\rm tr}\left[J^{N}_\mu \gamma_{\mu}^{T}\right].
\end{align}
\end{subequations}
These expressions readily yield the nucleon Dirac and Pauli form factors via Eq.\,\eqref{Sachs}.

It is worth noting that the results described herein are made with reference to a hadron scale, $\zeta_{\cal H}$, at which all properties of the subject hadron are carried by its dressed valence degrees of freedom \cite{Yin:2023dbw}.
Flavour-separa\-ted currents may therefore be defined as follows:
\begin{subequations}
\label{FlavourSep}
\begin{align}
\mathrm{u\, in\,proton:}\quad  &
2 J^{p }_{\mu} + J^{n  }_{\mu}
=3 [J_{00}^{(3)}]_\mu + [J_{11}^{(3)}]_\mu \,, \\
\mathrm{d\, in\,proton:}\quad  & J^{p }_{\mu} + 2 J^{n  }_{\mu}
= 2 [J_{11}^{(3)}]_\mu \,.
\end{align}
\end{subequations}
It follows that the in-proton $d$-quark form factors do not receive any contribution from MA components of the proton Faddeev amplitude.
This feature has a long-known analogue in the $q(qq)$ picture of nucleon structure; namely, hard probes do not perceive $d$ quarks unless axialvector diquark correlations are pre\-sent in the proton Faddeev wave function \cite{Close:1988br, Cui:2021gzg, Cheng:2023kmt}.
Using Eq.\,\eqref{Sachs}, flavour-separated Sachs form factors may be obtained from Eq.\,\eqref{FlavourSep}.

Canonical normalisation of the nucleon Faddeev amplitude is fixed by rescaling the amplitudes in Table~\ref{nucleonresults} so that one obtains $G_E^p(0)=1$.
Our formulation of the interaction current guarantees current conservation, which is expressed, \emph{e.g}., in the result $G_E^{n}(0)\equiv0$.

\begin{table}[t]
\caption{\label{TabcfExpt}
Nucleon magnetic moments (nuclear magnetons) and radii-squared (in fm$^2$), calculated using %conventional definitions -- Eq.\,\eqref{StandardDef}\,--\,supplementary material.
standard definitions, Eq.\,\eqref{StandardDef}.
The values in the column labelled ``$3$-body R'' are reproduced from Ref.\,\cite{Yao:2024uej}, which used a QCD-connected interaction in an analogous analysis.
Empirical values from Ref.\,\cite[PDG]{ParticleDataGroup:2024cfk}.
}
\begin{center}
\begin{tabular*}%{|c|c|c|c|c|c|c|}\hline
{\hsize}
{
l@{\extracolsep{0ptplus1fil}}
|l@{\extracolsep{0ptplus1fil}}
l@{\extracolsep{0ptplus1fil}}
l@{\extracolsep{0ptplus1fil}}}\hline
 & herein & $3$-body R & Exp.  \\\hline
 $\mu_p\ $ & $\phantom{-}2.17\ $ & $\phantom{-}2.23\ $ &$\phantom{-}2.793\ $  \\
 $\mu_n\ $ & $- 1.05\ $  & $-1.33\ $ & $-1.913\ $ \\
 $\langle r_E^2 \rangle^p$ & $\phantom{-}0.299\ $  & $\phantom{-}0.788\ $ & $\phantom{-}0.7070(7)\ $ \\
 $\langle r_E^2 \rangle^n$ & $-0.0722$ & $-0.0621\ $ & $-0.1160(22)\ $  \\
 $\langle r_M^2 \rangle^p$ & $\phantom{-}0.227\ $ & $\phantom{-}0.672\ $ & $\phantom{-}0.72(4)\ $ \\
 $\langle r_M^2 \rangle^n$ & $\phantom{-}0.220\ $  & $\phantom{-}0.661\ $ & $\phantom{-}0.75(2)\ $
 \\\hline
\end{tabular*}
\end{center}
\end{table}

\section{Nucleon Form Factors: Results}
\label{SecFFs}
In now providing comparisons between SCI form factor results and predictions obtained in realistic-interaction Faddeev equation studies \cite{Yao:2024uej}, both obtained using the same, sound CSM truncation, we enable identification of observables which are sensitive to the pointwise behaviour of the quark-quark interaction.
In this way, we highlight those experimental measurements which have the potential to chart that behaviour and distinguish between competing interaction pictures.
This is one of the principal purposes of SCI analyses and, as we shall see, electromagnetic form factors serve as valuable discerning tools.
\subsection{Sachs}
SCI predictions for nucleon static (low $Q^2$) properties are recorded in Table~\ref{TabcfExpt}.
In magnitude, the magnetic moments are $\sim 30$\% too small.
Similar magnetic moment values were reported in Ref.\,\cite{Yao:2024uej} -- $3$-body~R, which used a QCD-connected interaction to define the Faddeev equation and interaction current.
One may therefore conclude that underestimates of the size of nucleon magnetic moments is a failing of RL truncation.
This has previously been attributed to a flaw in the RL photon + quark vertex, whose dressed-quark anomalous magnetic moment term is too weak.
This weakness is corrected in higher-order truncations of the gap equation \cite{Chang:2010hb}.
Such corrections have been used in studies of mesons \cite{Xu:2022kng} and it may be possible to adapt that scheme to baryons.

\begin{figure}[t]
\vspace*{0.4em}

\leftline{\hspace*{0.5em}{\large{\textsf{A}}}}
\vspace*{-2ex}
\includegraphics[width=0.41\textwidth]{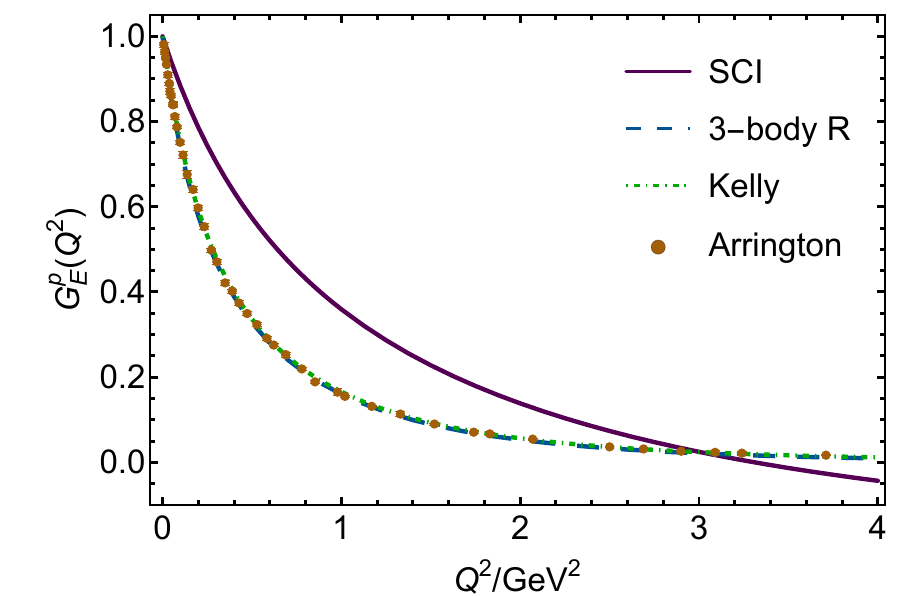}
\vspace*{0.1ex}
\leftline{\hspace*{0.5em}{\large{\textsf{B}}}}
\vspace*{-2ex}
\includegraphics[width=0.41\textwidth]{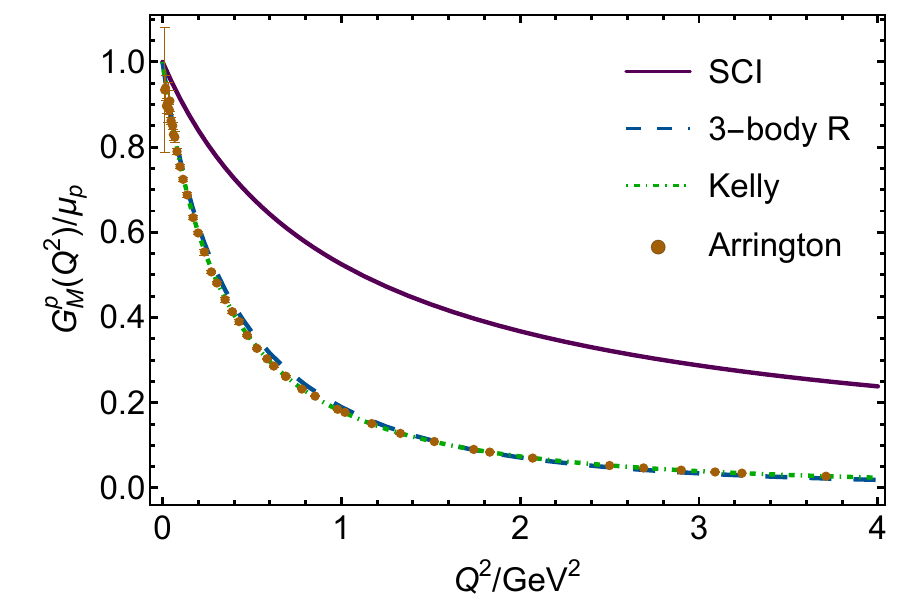}
\vspace*{3.5ex}

%%
%%\centering
%%\includegraphics[width=0.41\textwidth]{F6appA.pdf}\\
%%\includegraphics[width=0.41\textwidth]{F6appB.pdf}
%%\centering
\caption{\label{figGEMnpA}
Proton electromagnetic form factors.
Legend.
Solid purple curve -- SCI results obtained herein;
dashed blue curve  -- realistic-interaction $3$-body prediction, reproduced from Ref.\,\cite{Yao:2024uej};
dot-dashed green -- data parametrisation in Ref.\,\cite[Kelly]{Kelly:2004hm}.
Orange points -- experimental data taken from compilation in Ref.\,\cite{Arrington:2007ux}.}
%$G^p_E$ experimental data: Refs.\, Refs.~\cite{Punjabi:2005wq, Crawford:2006rz, Paolone:2010qc, JeffersonLabHallA:2011yyi, Zhan:2011ji, Arrington:2007ux}
%$G^p_M$ data: Ref.~\cite{Arrington:2007ux}.}
%; dashed, red - parametrisation of experimental
\end{figure}

Regarding the other entries in Table~\ref{TabcfExpt}, one sees that, as with mesons \cite{Roberts:2011wy, Chen:2012txa}, the SCI delivers radii that are typically too small in size, \emph{i.e}., hadrons that are too pointlike.
This is a readily anticipated consequence of using a hard interaction to define all Schwinger functions that enter into the calculation, \emph{viz}.\
a momentum-indepen\-dent interaction produces a momentum-indepen\-dent bound-state amplitude; hence, one must typically find that the associated bound-state form factors are too hard.
As highlighted by a comparison with Column~2, this SCI shortcoming is immediately remedied by using a QCD-connected quark-quark interaction.
Notwithstanding, the SCI $3$-body analysis does reproduce the ordering $\langle r_E^2 \rangle^p > \langle r_M^2 \rangle^p$ found via contemporary analyses of existing form factor measurements \cite{Cui:2022fyr}.

\begin{figure}[t]
\vspace*{0.4em}

\leftline{\hspace*{0.5em}{\large{\textsf{A}}}}
\vspace*{-2ex}
\includegraphics[width=0.41\textwidth]{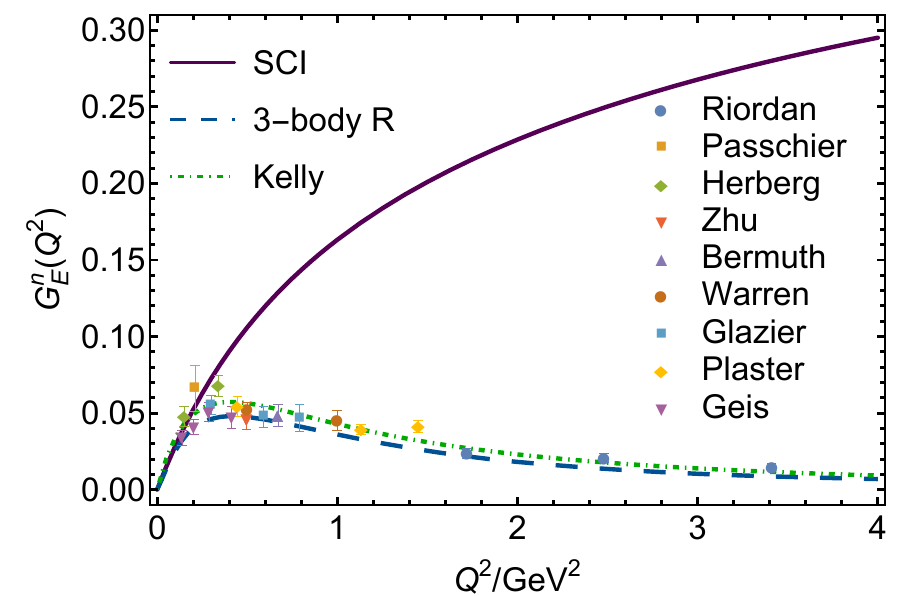}
\vspace*{0.1ex}
\leftline{\hspace*{0.5em}{\large{\textsf{B}}}}
\vspace*{-2ex}
\includegraphics[width=0.41\textwidth]{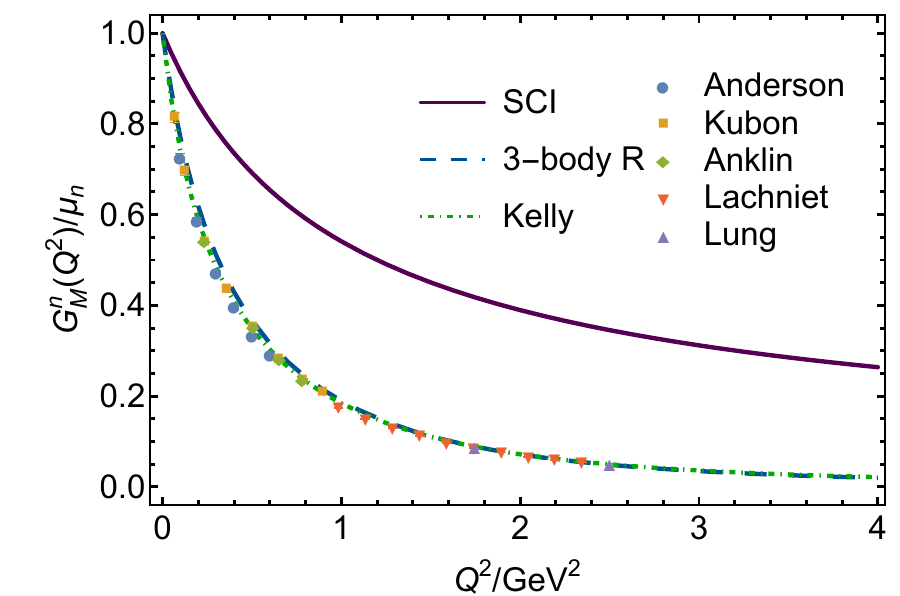}
\vspace*{3.5ex}

%%
%%\centering
%%\includegraphics[width=0.41\textwidth]{F6appA.pdf}\\
%%\includegraphics[width=0.41\textwidth]{F6appB.pdf}
%%\centering
\caption{\label{figGEMnpB}
Neutron electromagnetic form factors.
Legend.
Solid purple curve -- SCI results obtained herein;
dashed blue curve  -- realistic-interaction $3$-body prediction, reproduced from Ref.\,\cite{Yao:2024uej};
dot-dashed green -- data parametrisation in Ref.\,\cite[Kelly]{Kelly:2004hm}.
$G^n_E$ experimental data: Refs.\,\cite{Passchier:1999cj, Herberg:1999ud, E93026:2001css, Bermuth:2003qh, Warren:2003ma, Glazier:2004ny, Plaster:2005cx, BLAST:2008bub, Riordan:2010id}.
$G^n_M$ data: Refs.\,\cite{Lung:1992bu, Anklin:1998ae, Kubon:2001rj, JeffersonLabE95-001:2006dax, CLAS:2008idi}}
%
%Experimental data for $G^n_E$ and $G^n_M$ are, respectively, from Refs.\,\cite{Rock:1982gf, Lung:1992bu, Anklin:1998ae, Kubon:2001rj, JeffersonLabE95-001:2006dax, CLAS:2008idi, BLAST:2008bub, Passchier:1999cj, Herberg:1999ud, Riordan:2010id, E93026:2001css, Bermuth:2003qh, Warren:2003ma, Glazier:2004ny, Plaster:2005cx} and Refs.\,\cite{Rock:1982gf, Lung:1992bu, Anklin:1998ae, Kubon:2001rj, JeffersonLabE95-001:2006dax, CLAS:2008idi}.}%; dashed, red - parametrisation of experimental data~\cite{Kelly:2004hm}
\end{figure}

Turning to Figs.\,\ref{figGEMnpA}, \ref{figGEMnpB}, we record that here and hereafter all plots are restricted to the spacelike domain $Q^2\leq 4\,$GeV$^2\approx 5\,\Lambda_{\rm uv}^2$.  The SCI cannot be realistic on the complement of this domain, whose boundary is already much larger than the interaction's ultraviolet cutoff.
This is highlighted, \emph{e.g}., by the fact that, again, that SCI Faddeev equation predictions for the overall $Q^2$ dependence of each nucleon form factor is too hard.
On the other hand, the $3$-body~R calculation delivers results in agreement with data
\cite{Arrington:2007ux, Passchier:1999cj, Herberg:1999ud, E93026:2001css, Bermuth:2003qh, Warren:2003ma, Glazier:2004ny, Plaster:2005cx, BLAST:2008bub, Riordan:2010id, Lung:1992bu, Anklin:1998ae, Kubon:2001rj, JeffersonLabE95-001:2006dax, CLAS:2008idi}.
%(See Figs.\,\ref{figGEMnpA}, \ref{figGEMnpB}\,--\,supplementary material for confirmation.)
%(See Figs.\,S.1, S.2\,--\,supplementary material for confirmation.)

Notably, even the SCI delivers $G_E^n(Q^2)\not\equiv 0$ -- see Fig.\,\ref{figGEMnpB}\,A, an outcome which highlights that a Poincar\'e-covariant treatment of quark-quark interactions necessarily delivers nucleon wave functions that do not possess SU$(4)$ flavour-spin symmetry, \emph{viz}.\ $u$ and $d$ quark wave functions that are not identical, even after removing simple counting factors.
This also has implications for nucleon structure functions; see, \emph{e.g}., Refs.\,\cite{Chang:2022jri, Lu:2022cjx}.
%% is a fair match, despite its sensitivity to details of the neutron wave function, especially as expressed in $F_1^n$ -- see, \emph{e.g}., Refs.\,\cite{Eichmann:2016yit, Segovia:2014aza}.
%Notably, despite underestimating $\mu_{p,n}$, the $Q^2$-dependences of the proton and neutron magnetic form factors agree well with experiment.

\subsection{Form Factor Ratios}
\label{NFFR}
\begin{figure}[t]
\vspace*{1.2em}

\leftline{\hspace*{0.5em}{\large{\textsf{A}}}}
\vspace*{-5ex}
\includegraphics[clip, width=0.41\textwidth]{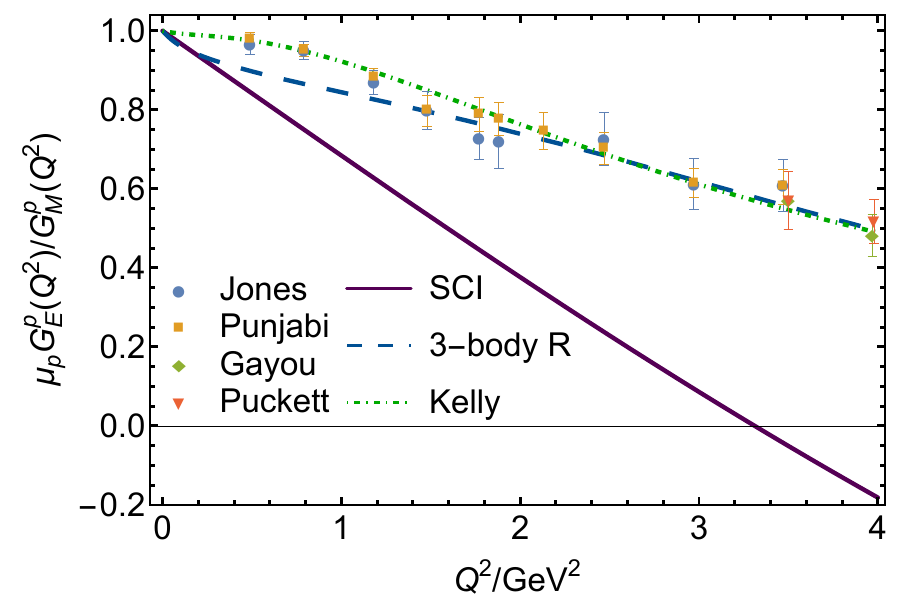}
\vspace*{1ex}
\leftline{\hspace*{0.5em}{\large{\textsf{B}}}}
\vspace*{-5ex}
\includegraphics[clip, width=0.41\textwidth]{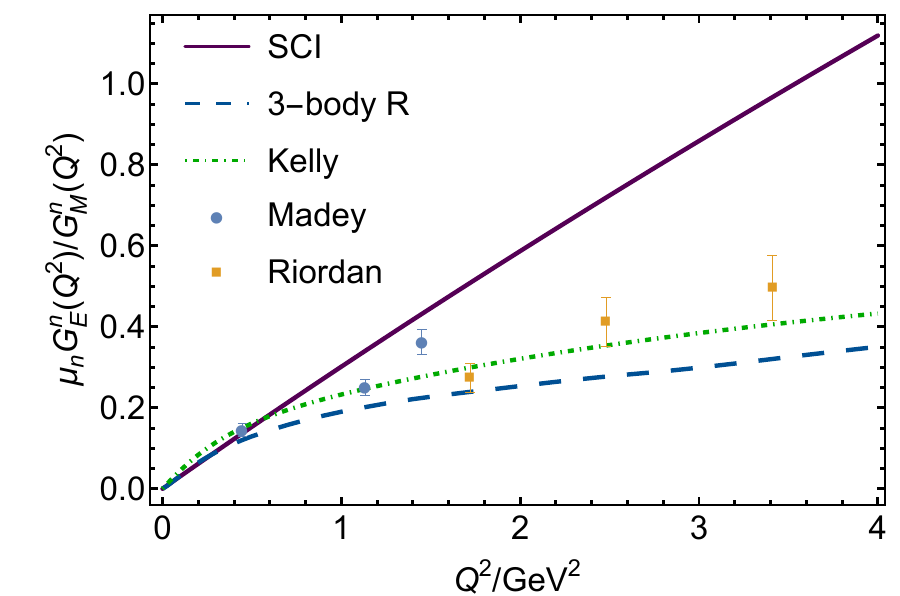}
\vspace*{4.5ex}
\caption{\label{mupGEpGMp}
%
%{\sf Panel A} $\mu_p G_E^p(Q^2)/G_M^p(Q^2)$.
{\sf Panel A}: $\mu_p G_E^p/G_M^p$.
%{\sf Panel B} $\mu_n G_E^n(Q^2)/G_M^n(Q^2)$.
{\sf Panel B}: $\mu_n G_E^n/G_M^n$.
Legend.
Solid purple curve -- SCI results obtained herein;
dashed blue curve -- realistic-interaction $3$-body prediction, reproduced from Ref.\,\cite{Yao:2024uej};
dot-dashed green -- data parametrisation in Ref.\,\cite[Kelly]{Kelly:2004hm}.
Data: proton -- Refs.\,\cite{Jones:1999rz, Gayou:2001qd, Punjabi:2005wq, Puckett:2010ac, Puckett:2017flj}; and neutron -- Refs.\,\cite{Madey:2003av, Riordan:2010id}.
%(Jones~\cite{JeffersonLabHallA:1999epl},  Gayou~\cite{JeffersonLabHallA:2001qqe}, Punjabi~\cite{Punjabi:2005wq}, Puckett(2010)~\cite{Puckett:2010ac}, and Puckett(2011)~\cite{Puckett:2011xg})
%
}
\end{figure}

Since the pioneering polarisation transfer experiment described in Ref.\,\cite{Jones:1999rz}, the form factor ratios\linebreak $\mu_N G_E^N(Q^2)/G_M^N(Q^2)$ have been of great interest.  SCI predictions are drawn in Fig.\,\ref{mupGEpGMp} and compared therein with kindred calculations and data.

The SCI predicts a zero in $\mu_p G_E^p(Q^2)/G_M^p(Q^2)$ at $Q^2=3.25 {\rm{GeV}}^2$.
This location is similar to that found in $q(qq)$ treatments of the proton \cite[Fig.\,7]{Wilson:2011aa}.
As observed therein and made plain by a comparison between SCI and $3$-body~R results in Fig.\,\ref{mupGEpGMp}\,A, the existence of a zero in this ratio is independent of the quark-quark interaction used in formulating the Poincar\'e-covariant nucleon bound-state problem.
The location, on the other hand, is a sensitive measure of the interaction used and, therefore, a signature of EHM and gauge sector dynamics.

It is worth noting here that Ref.\,\cite{Cheng:2024cxk}, which subjected the data reproduced in Fig.\,\ref{mupGEpGMp}\,A to an objective, model-independent analysis, reached the following conclusions:
with 50\% confidence, extant $\mu_p G_E^p(Q^2)/ G_M^p(Q^2)$ polarisation transfer data are consistent with the existence of a zero in the ratio on $Q^2 \leq 10.37\,$GeV$^2$;
the level of confidence rises to 99.9\% on $Q^2 \leq 13.06\,$GeV$^2$;
and the likelihood that the data are consistent with the absence of a zero in the ratio on $Q^2 \leq 14.49\,$GeV$^2$ is $1/1$-million.
The earlier, hence independent, $3$-body~R treatment of the problem \cite{Yao:2024uej} found a zero at \begin{equation}
\label{ProtonZero}
Q^2 /{\rm GeV}^2= 8.86^{+1.93}_{-0.86}.
\end{equation}
This value falls easily within the domain determined empirically in Ref.\,\cite{Cheng:2024cxk}.

The SCI result for $\mu_n G_E^n(Q^2)/G_M^n(Q^2)$ is drawn in Fig.\,\ref{mupGEpGMp}\,B, along with kindred calculations and data.
Like the $3$-body~R result \cite{Yao:2024uej}, the SCI predicts that this ratio rises uniformly with increasing $Q^2$.  Naturally, the SCI result is too stiff.
Like the $3$-body~R study, therefore, the SCI predicts that there is a $Q^2$ domain upon which the charge form factor of the neutral neutron is larger than that of the positively charged proton.
Using the SCI, this domain begins at $Q^2 = 1.6\,$GeV$^2$, whereas that point lies at $Q^2 = 4.66 ^{+0.18}_{-0.13}$GeV$^2$ in the $3$-body~R approach.
%% 4.66+0.18-0.13

\subsection{Form Factor Flavour Separation}
\label{NFFFS}
Working from Eq.\,\eqref{FlavourSep}, one obtains the following expressions for the hadron-scale flavour-separated in-pro\-ton Dirac and Pauli form factors:
\begin{equation}
\label{FlavourSep}
F_{i}^u = 2 F_{i}^p + F_{i}^{n}, \;
F_{i}^d = F_{i}^{p} + 2 F_{i}^{n}, \; i=1,2\,.
\end{equation}
Current conservation and valence-quark number entail
%%\begin{equation}
%%\label{FlavourNorm}
$F_{1}^{u}(Q^2=0)=2= 2 F_{1}^{d}(Q^2=0)$.
%%\end{equation}

\begin{figure}[t]
\vspace*{1.2em}

\leftline{\hspace*{0.5em}{\large{\textsf{A}}}}
\vspace*{-5ex}
\includegraphics[clip, width=0.41\textwidth]{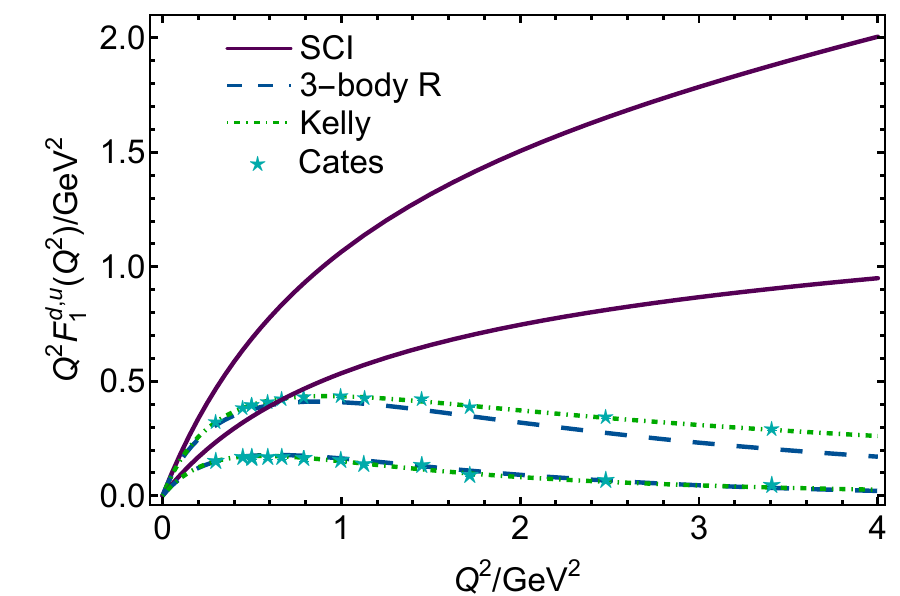}
\vspace*{1ex}
\leftline{\hspace*{0.5em}{\large{\textsf{B}}}}
\vspace*{-5ex}
\includegraphics[clip, width=0.41\textwidth]{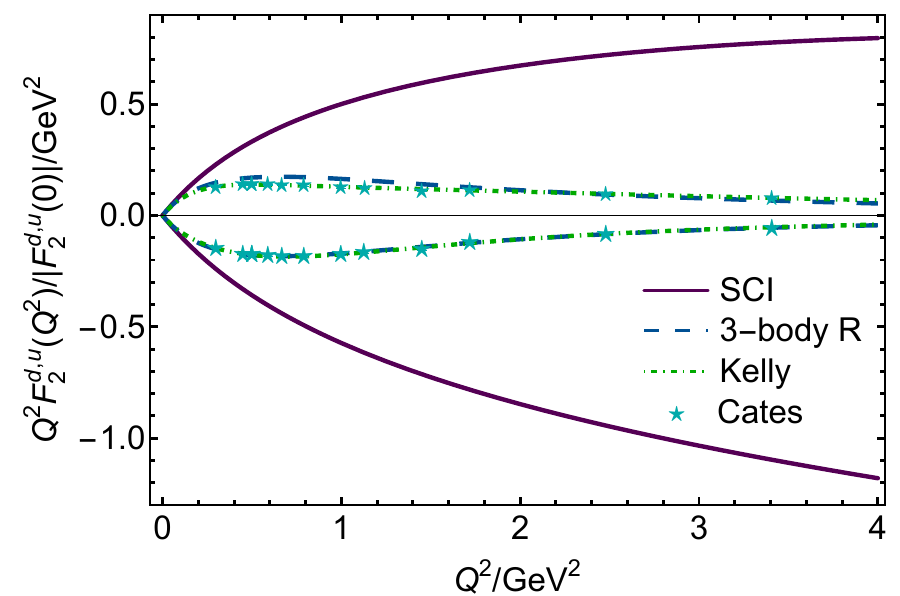}
\vspace*{4.5ex}
\caption{\label{FlavSep}
Flavour-separated proton form factors (hadron scale): $Q^2 F_1^{d,u}(Q^2)$ ({\sf Panel A}); and $Q^2 F_2^{d,u}(Q^2)/|F_2^{d,u}(0)/|$ ({\sf Panel B}).
Legend.
Solid purple curve -- SCI results obtained herein;
dashed blue curve  -- realistic-interaction $3$-body prediction, reproduced from Ref.\,\cite{Yao:2024uej};
dot-dashed green -- data parametrisation in Ref.\,\cite[Kelly]{Kelly:2004hm}.
Data -- cyan 5-pointed stars \cite{Cates:2011pz}.
In all cases, each $d$ quark result is smaller than that of the $u$ quark.
}
\end{figure}

SCI results for these form factors are drawn in Fig.\,\ref{FlavSep}.
To account for the RL truncation underestimate of nucleon magnetic moments, both theory and experiment Pauli form factors in Fig.\,\ref{FlavSep}\,B are normalised by the magnitude of their $Q^2=0$ values.
Again, for the reasons explained above, the SCI predictions are too stiff; and the differences between the SCI and $3$-body~R results reveal the very significant influence on observables of momentum-dependent gluon and quark running masses.
Notably, the $3$-body~R predictions are in good agreement with available data.

Importantly, the $3$-body SCI predicts no zero in $F_1^d$ on $Q^2 \lesssim 10\,$GeV$^2$.
On the other hand, the $q(qq)$ SCI treatment of this problem in Ref.\,\cite{Wilson:2011aa} delivers a zero at $Q^2 \approx 10\,\Lambda_{\rm ir}^2$; as this is far above the anticipated domain of SCI applicability, one needs to treat the result with caution.
Notwithstanding that, the $3$-body~R approach also predicts a zero in $F_1^d$ at
\begin{equation}
Q^2_{F_1^d{\rm -zero}} =5.73^{+1.46}_{-0.49}\,{\rm GeV}^2.
\end{equation}
This outcome matches the result obtained in a realistic-interaction $q(qq)$ treatment of the nucleon \cite{Cheng:2025yij}: $Q^2=5.1^{+0.2}_{-0.1}\,$GeV$^2$.

Expanding upon these remarks, it is worth stressing that neither the $3$-body SCI nor the $3$-body~R treatment of the nucleon predict a zero in any other form factor that appears in the panels of Fig.\,\ref{FlavSep}.

\begin{figure}[t]
\centering
\includegraphics[width=0.41\textwidth]{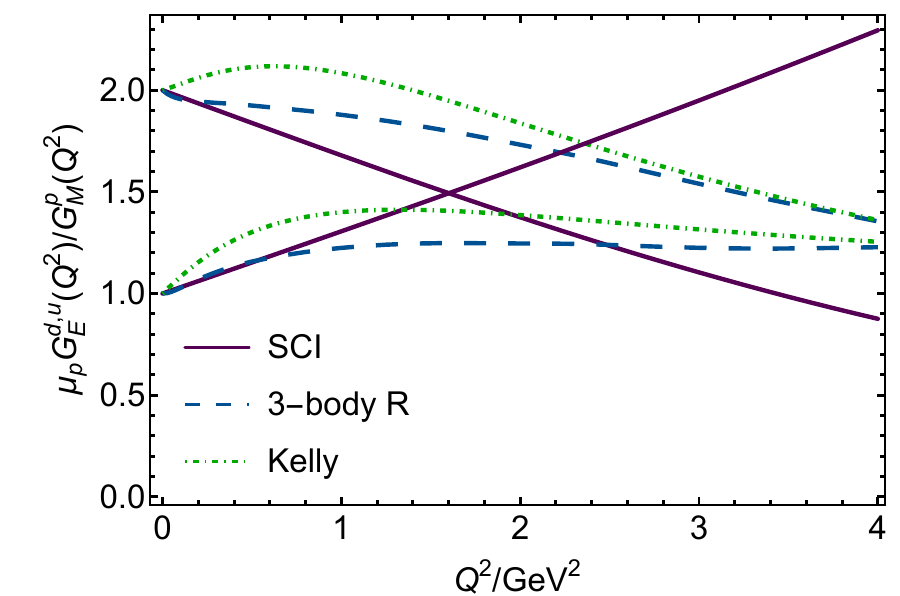}
\caption{\label{FigFlavSepGEM}
Flavour separation of charge and magnetisation form factors (hadron scale), with each function normalised by $G_M^p$ so as to highlight their differing $Q^2$-dependence.
Legend.
Solid purple curve -- SCI results obtained herein;
dashed blue curve  -- realistic-interaction $3$-body prediction, reproduced from Ref.\,\cite{Yao:2024uej};
dot-dashed green -- data parametrisation in Ref.\,\cite[Kelly]{Kelly:2004hm}.
Data -- cyan 5-pointed stars \cite{Cates:2011pz}.
The $d$ quark results are unity at $Q^2=0$, whereas those for the $u$ quark take the value $2$.}
%
%Experimental data for $G^n_E$ and $G^n_M$ are, respectively, from Refs.\,\cite{Rock:1982gf, Lung:1992bu, Anklin:1998ae, Kubon:2001rj, JeffersonLabE95-001:2006dax, CLAS:2008idi, BLAST:2008bub, Passchier:1999cj, Herberg:1999ud, Riordan:2010id, E93026:2001css, Bermuth:2003qh, Warren:2003ma, Glazier:2004ny, Plaster:2005cx} and Refs.\,\cite{Rock:1982gf, Lung:1992bu, Anklin:1998ae, Kubon:2001rj, JeffersonLabE95-001:2006dax, CLAS:2008idi}.}%; dashed, red - parametrisation of experimental data~\cite{Kelly:2004hm}
\end{figure}

Hadron-scale flavour separations of the Sachs form factors are obtained using ($e_u=2/3$, $e_d=-1/3$):
\begin{equation}
\label{FlavSepGEM}
G_E^p  = e_u G_E^{pu} + e_d G_E^{pd}\,, \quad
G_E^n  = e_u G_E^{pd} + e_d G_E^{pu}\,.
\end{equation}
SCI results for these form factors are drawn in Fig.\,\ref{FigFlavSepGEM}, wherein the are compared with $3$-body~R results.
These images elucidate that $G_E^p$ exhibits a zero because, although it is non-negative, $G_E^{pu}/G_M^p$ falls steadily with increasing $Q^2$ whereas $G_E^{pd}/G_M^p$ is positive and increasing.
Conversely and consequently, $G_E^n$ does not possess a zero because $e_u>0$, $G_E^{pd}/G_M^p$ is large, positive and increasing, and $|e_d G_E^{pu}|$ is always less than $e_u G_E^{pd}$.
Similar arguments hold for the $3$-body~R treatment, except for the modification that $G_E^{pd}/G_M^p$ is positive and approximately constant.  This difference shifts the zero in $G_E^p$ to a larger value of $Q^2$ relative to that found with the SCI; see Eq.\,\eqref{ProtonZero}.

The character of $G_E^{pd}/G_M^p$ owes to the fact that $F_2^d$ is negative definite and decreasing on the entire domain displayed in Fig.\,\ref{FlavSep} and $F_1^d$ is positive and steadily increasing; so, $G_E^d=F_1^d - (Q^2/[4 m_N]^2]) F_2^d$ must increase steadily.
The ratio $G_E^{pu}/G_M^p$,
$G_E^u=F_1^u - (Q^2/[4 m_N]^2]) F_2^u$,
falls steadily with increasing $Q^2$ because $F_1^u$ is monotonically decreasing and $Q^2F_2^u$ is monotonically increasing.
Appropriately modified for the different behaviour of $F_1^d$, analogous arguments can be made in connection with the $3$-body~R results.

It is worth recalling the arguments made in closing Ref.\,\cite[Sect.\,3]{Yao:2024uej}, which establish that it is normal for the electric form factor of an electrically charged $J\neq 0$ bound state to possess a zero because of the potential for destructive interference between the leading charge form factor and magnetic and higher multipole form factors; see, \emph{e.g}., Eq.\,\eqref{Sachs}.
This is not true for $J=0$ \cite{Yao:2024drm}: such systems have only one electromagnetic form factor, $F_{J=0}$, and both valence contributions to $F_{J=0}$ have the same sign.

\section{Summary and Perspective}
\label{Epilogue}
Working with a symmetry-pre\-ser\-ving treatment of a vector\,$\otimes$\,vector contact interaction (SCI), we introduced a largely algebraic $3$-body Faddeev equation treatment of the nucleon bound state problem and used it to deliver results for
all nucleon charge and magnetisation distributions and their flavour separation.
A significant merit of the SCI treatment is that it provides for a transparent understanding of the $3$-body Faddeev equation and its use in developing predictions for baryon observables.

This is the first SCI treatment of the nucleon as a $3$-body problem; so it is worth recapitulating some of the outcomes herein that are either novel or complementary to existing treatments of the nucleon using continuum Schwinger function methods (CSMs).

In a $3$-body treatment of the nucleon with a QCD-connected quark-quark interaction ($3$-body\,R), $64$ independent components are required to completely specify the system's Faddeev amplitude.
Using the SCI, all components of the nucleon Faddeev amplitude that are associated with quark + quark relative momentum vanish.  This reduces the number of independent components to $8$ [Sect.\,\ref{SubSecFE}].
In the momentum-dependent quark + diquark treatment ($q(qq)$\,R), the Faddeev amplitude also has $8$ independent components, but they are very different: two describe the isoscalar-scalar diquark piece and six are associated with the isovector axialvector part, and all eight are momentum-dependent.
%\
Furthermore, with isospin symmetry assumed, the $3$-body amplitude possesses $S_3$ symmetry.  Exploiting this [Sect.\,\ref{S3FA}], the SCI amplitude reduces to just $3$ independent terms because, \emph{e.g}., the mixed-symmetric (MS) flavour components are fully determined by the mixed-antisymmetric (MA) terms.
The standard SCI $q(qq)$ treatment of the nucleon does not possess $S_3$ symmetry because the isoscalar-scalar and isovector-axialvector diquark correlations are not mass-degenerate; nevertheless, the associated Faddeev amplitude also has only $3$ components: $8\to 3$ because the amplitude cannot depend on $q$ + $(qq)$ relative momentum.

Given the above observations, it is plain that comparisons between the electromagnetic form factors obtained with the four complementary approaches can provide insights into the roles played in nucleon structure by dynamics and symmetries. %orbital angular momentum and the presence of tight diquark correlations.
For instance, comparing $3$-body SCI predictions with those obtained using $3$-body\,R [Sect.\,\ref{SecFFs}], one finds that $Q^2\simeq 0$ results are similar and there are many qualitative correspondences between $Q^2$-dependent form factor properties, \emph{e.g}.,
the proton electric form factor possesses a zero and the neutron electric form factor is nonzero.
These outcomes are an expression of the underlying Poincar\'e-covariant Faddeev equation.
However, the spacelike $Q^2$ evolution of SCI form factors is typically too stiff: this is unsurprising, given that the interaction itself is hard.
Similar statements can be made concerning the contrast between SCI $q(qq)$ studies and $q(qq)$\,R results.
A clear distinction, between SCI $3$-body and $q(qq)$ analyses is found in the behaviour of the $d$-quark-in-proton Dirac form factor: whereas SCI $q(qq)$ produces a zero in $F_1^d$, SCI $3$-body does not.

These points illustrate that in combining the SCI $3$-body predictions herein with those already obtained for nucleon properties using the complementary CSM formulations, one arrives at a broad view of the impacts of, \emph{inter alia}, Poincar\'e-covariance, the pointwise behaviour of the quark-quark interaction, orbital angular momentum within the nucleon, and phenomena associated with the emergence of hadron mass.

Numerous extensions of this work are possible.
For instance, a first step would be to treat the entire array of SU$(3)$ octet and decuplet baryons in the same way.
This could potentially provide a novel and simple means of understanding the effects of SU$(3)$ symmetry breaking on baryon observables, like the $\Lambda - \Sigma^0$ mass splitting, which appears naturally in modern quark + diquark, $q(qq)$, descriptions of baryon structure \cite{Chen:2019fzn, Yin:2019bxe, Cheng:2022jxe}.  Such a study is underway.
Its successful completion may enable a further extension of our framework to baryons containing one or more heavy quarks.  In such cases, a sound treatment of the huge disparity between emergent and explicit (Higgs-boson-generated) mass-scales and its expression in observables poses a significant challenge to all extant predictive frameworks.

An SCI treatment of baryon semileptonic weak transition form factors would also be valuable.
Regarding the form factors involved, SCI $q(qq)$ analyses either already exist \cite{Cheng:2022jxe} or are underway.
Their comparison with results from an SCI $3$-body study may provide clear insights into differences introduced by assuming the existence of tight diquark correlations as opposed to having loose correlations emerge dynamically.
Such a study would also provide a useful guide to and benchmark for realistic-interaction $3$-body treatments of semileptonic transitions.
As with analogous decays involving mesons, these processes open a window onto interference between the impacts of Higgs-boson couplings into QCD and emergent hadron mass (EHM) \cite{Roberts:2021nhw, Ding:2022ows}: the Higgs produces quark current masses, but EHM is the key to explaining dressed quark masses, which are $\sim 100$ times larger.

One might also use the SCI $3$-body treatment to describe nucleon $\to$ resonance electroweak transitions \cite{Achenbach:2025kfx}, for which, thus far, Poincar\'e-covariant analyses that employ continuum Schwinger function methods have largely been based on the $q(qq)$ picture; see, \emph{e.g}., Refs.\,\cite{Eichmann:2011aa, Segovia:2014aza, Lu:2019bjs, Raya:2021pyr, Chen:2023zhh}.  An exception is Ref.\,\cite{Sanchis-Alepuz:2017mir}; and in this connection, too, comparisons with SCI $3$-body results could prove instructive.

%\acknowledgments
%\noindent\emph{Acknowledgments}\,---\,%
\begin{acknowledgements}
We are grateful for constructive interactions with D.\ Binosi and for his assistance in preparing Figs.\,1\,-\,3.
% S.-X.\ Qin.
%
Work supported by:
National Natural Science Foundation of China (grant nos.\ 12135007, 12205149);
and
Helmholtz-Zentrum Dresden-Rossendorf, under the High Potential
Programme.

\medskip

\noindent\textbf{Data Availability Statement} Data will be made available on reasonable request.  [Authors' comment: All information necessary to reproduce the results described herein is contained in the material presented above.]
\medskip

\noindent\textbf{Code Availability Statement} Code/software will be made available
on reasonable request. [Authors' comment: No additional remarks.]

\end{acknowledgements}

\appendix

\section{Regularised Integrals}
\label{AppendixA}
\subsection{One loop}
\label{AppendixA1}
In SCI studies, numerous divergent integrals arise.
We give them meaning by implementing the scheme discussed below \cite{Xing:2022jtt}.
This variant is practically equivalent to that introduced in Ref.\,\cite{Gutierrez-Guerrero:2010waf}; but it leads to somewhat simpler ``bookkeeping'', so can be easier to implement mechanically.

First consider:
\begin{subequations} %%% multiply by 16 Pi^2 ...
\label{CnIntegrals}
\begin{align}
{\mathsf C}_{-2\alpha}(\sigma) & = \int \frac{d^4 l}{\pi^2} \frac{1}{[l^2+\sigma]^{2+\alpha}}\\
& = \int \frac{d^4 l}{\pi^2} \int_0^{\infty} d\tau \frac{\tau^{1+\alpha}}{\Gamma(2+\alpha)} {\rm e}^{-\tau [l^2+\sigma]}\\
& = \int_0^\infty d\tau
\frac{\tau^{\alpha-1}}{\Gamma(2+\alpha)} {\rm e}^{-\tau \sigma}\,.
\end{align}
\end{subequations}
Each of these steps is rigorously well defined for $\alpha>0$.   In SCI applications, however, one typically encounters cases with $\alpha=-1, 0$, in which case the integrals are quadratically or logarithmically divergent, respectively.
We therefore follow Ref.\,\cite{Ebert:1996vx} and introduce the following proper-time regularisation of all integrals:
\begin{subequations}
\begin{align}
& {\mathsf C}_{-2\alpha}^{\rm iu} (\sigma)
= \int_{\tau_{\rm uv}}^{\tau_{\rm ir}} d\tau
\frac{\tau^{\alpha-1}}{\Gamma(2+\alpha)} {\rm e}^{-\tau \sigma} \\
& = \frac{1}{\Gamma(2+\alpha)}  \frac{1}{\sigma^\alpha}
[
\Gamma(\alpha, \tau_{\rm uv}\sigma)
-\Gamma(\alpha, \tau_{\rm ir}\sigma) ] \\
& = \frac{1}{\sigma^\alpha} \overline{\cal C}^{\rm iu}_{\alpha+1}(\sigma)\,, \label{OldRegNew}
%
%%\frac{1}{\tau_{\rm uv}^{2-\alpha}} {\rm E}_{\alpha-1}(\tau_{\rm uv}\sigma)
% -\frac{1}{\tau_{\rm ir}^{2-\alpha}} {\rm E}_{\alpha-1}(\tau_{\rm ir}\sigma)\bigg]\,,
\end{align}
\end{subequations}
where
$\Gamma(\alpha,y)$ is the incomplete gamma-function, \linebreak
$\overline{\cal C}^{\rm iu}_{\alpha+1}(\sigma)$ are the regularisation functions exploited in Ref.\,\cite{Gutierrez-Guerrero:2010waf},
%${\rm E}_{\mathpzc a}(z)$ is a generalised exponential-integral function
and $\tau_{\rm ir}=1/\Lambda_{\rm ir}^2$, $\tau_{\rm uv}=1/\Lambda_{\rm uv}^2$ are, respectively, infrared and ultraviolet regulator scales.
A nonzero value of $\Lambda_{\rm ir}$ implements confinement by eliminating quark production thresholds
\cite[Sect.\,5]{Ding:2022ows}.
%\cite{Roberts:2007ji}.
A nonzero value of $\tau_{\rm uv}$ is necessary to produce a finite result for the integral; hence, the value of $\Lambda_{\rm uv}$ becomes a dynamical scale in the SCI and, therefore, a part of its definition: it sets the scale for all mass-dimensioned quantities and, loosely, an upper bound on the domain of SCI validity.

Another example is
\begin{subequations}
\begin{align}
{\mathsf D}_{-2\alpha}&(\sigma)  \delta_{\mu\nu} =
\int \frac{d^4 l}{\pi^2} \frac{l_\mu l_\nu}{[l^2+\sigma]^{3+\alpha}}\\
& = \delta_{\mu\nu} \int \frac{d^4 l}{\pi^2} \frac{l^2}{4}
\int_0^\infty  d\tau \frac{\tau^{2+\alpha}}{\Gamma(3+\alpha)} {\rm e}^{-\tau [l^2+\sigma]}\\
& = \delta_{\mu\nu} \frac{1}{2}
 \int_0^\infty  d\tau  \frac{\tau^{\alpha-1}}{\Gamma(3+\alpha)} {\rm e}^{-\tau \sigma}\,.
\end{align}
\end{subequations}
In this connection, we introduce
%% Gamma[3+alpha] = Gamma[1+2+alpha] = (2+alpha) Gamma[2+alpha]
\begin{subequations}
\begin{align}
 {\mathsf D}_{-2\alpha}^{\rm iu} (\sigma)
& = \frac{1}{2} \int_{\tau_{\rm uv}}^{\tau_{\rm ir}} d\tau
\frac{\tau^{\alpha-1}}{\Gamma(3+\alpha)} {\rm e}^{-\tau \sigma} \\
& = \frac{1}{2} \frac{\Gamma(2+\alpha)}{\Gamma(3+\alpha)}
{\mathsf C}_{-2\alpha}^{\rm iu} (\sigma)\\
& = \frac{1}{2} \frac{1}{2+\alpha}
{\mathsf C}_{-2\alpha}^{\rm iu} (\sigma)\,.
\end{align}
\end{subequations}

Following the above procedures, one is expressing the following identity:
\begin{align}
& 4 {\mathsf D}_{-2\alpha}^{\rm iu} (\sigma) =
\int d^4 l \frac{l^2}{[l^2+\sigma]^{3+\alpha}} \nonumber \\
%
%& \stackrel{\rm (regularised)}{=}
& \stackrel{\rm reg.}{=}
\frac{2}{2+\alpha}
\int d^4 l \frac{1}{[l^2+\sigma]^{2+\alpha}} =
\frac{2}{2+\alpha}{\mathsf C}_{-2\alpha}^{\rm iu} (\sigma)\,.
\label{IntegralIdentities}
\end{align}
Of course, this is true for any value of $\alpha$ for which both integrals are finite.  Regularisation defines Eq.\,\eqref{IntegralIdentities} to be true, by analytic continuation, at all other $\alpha$ values.

Note that:
\begin{equation}
{\mathsf D}_{0}^{\rm iu} (\sigma) =  (1/4){\mathsf C}_{0}^{\rm iu} (\sigma)\,,
{\mathsf D}_{2}^{\rm iu} (\sigma) =  (1/2){\mathsf C}_{2}^{\rm iu} (\sigma)\,.
\end{equation}
%C2 - 2 D2
In ensuring, \emph{e.g}., the axialvector Ward-Green-Takahashi identity, the latter relation merely suggests a different rearrangement of terms in the integrand so as to ensure Ref.\,\cite[Eq.\,(17)]{Gutierrez-Guerrero:2010waf}, \emph{viz}.
\begin{align}
\int \frac{d^4 l}{\pi^2} \frac{\tfrac{1}{2} l^2 + \sigma}{[l^2 + \sigma]^2} &
\to \frac{1}{2} [ {\cal C}_0^{\rm ir}(\sigma) + {\cal C}_1^{\rm ir}(\sigma) ] \\
&
= \int \frac{d^4 l}{\pi^2}
\left[ \frac{1}{l^2 + \sigma} - \frac{1}{2} \frac{l^2}{[l^2+\sigma]^2}\right] \\
& \to {\mathsf C}_{2}^{\rm ir}(\sigma) - 2 {\mathsf D}_{2}^{\rm ir}(\sigma) = 0 \,.
%
%%& = \frac{1}{\sigma} \left[ {\cal C}_0^{\rm ir}(\sigma) \right]
\end{align}
This elucidates the connection between Eq.\,\eqref{IntegralIdentities} and an analogue in Ref.\,\cite{Gutierrez-Guerrero:2010waf}, \emph{i.e}.\
$\overline{\cal C}^{\rm iu}_{0}(\sigma) + \overline{\cal C}^{\rm iu}_{1}(\sigma) = 0$.

%\medskip

\subsection{Two loop: interaction current}
\label{AppendixA2}
In calculating form factors from the current sketched in Fig.\,\ref{FigCurrent}, one encounters two loop integrals with the following structure:
\begin{equation}
\mathpzc T(Q;P) = \int_{\sf dp} \int_{\sf dq} \frac{{\cal N}(p,q,P)}{{\cal D}(p,q,P)}
\end{equation}
where
\begin{align}
{\cal D}(p,q,P)
& = (p_1^2+M^2)(p_2^2+M^2) \nonumber \\
& \quad \times (p_3^{+}\cdot p_3^+ + M^2)(p_3^{-}\cdot p_3^-+M^2)
\end{align}
and the internal quark momenta are given in Eq.\,\eqref{p1p2p3}.

Now introduce a first Feynman parametrisation parameter, ${\mathpzc a}_1$, and shift the $q$ integration variable $q \to q -p/2 +{\mathpzc a}_1 p - P/3(1-2{\mathpzc a}_1)$  to obtain:
\begin{subequations}
\label{FP12L}
\begin{align}
& \frac{1}{(p_1^2+M^2)(p_2^2+M^2)}
= \int_0^1 d{\mathpzc a}_1 1/D_{12}({\mathpzc a}_1,q,p)^2\,, \\
& D_{12}({\mathpzc a}_1,q^2,p)  = q^2 + M^2  \nonumber \\
&\quad + {\mathpzc a}_1(1-{\mathpzc a}_1)(p^2 - (4/3) p\cdot P +(4/9) P^2) \,.
\end{align}
\end{subequations}
This integrand is an even function of $q$.

Next, a second Feynman parametrisation parameter, to write:
\begin{subequations}
\label{FP22L}
\begin{align}
& \frac{1}{ (p_3^{+}\cdot p_3^+ + M^2)(p_3^{-}\cdot p_3^- + M^2)}
= \int_0^1 d{\mathpzc a}_2 1/D_{3}({\mathpzc a}_2,p)^2\,, \\
& D_{3}({\mathpzc a}_2,p)  =p^2 + M^2  \nonumber \\
&\quad (1 - 2 {\mathpzc a}_2) p\cdot Q + (2/3) p\cdot P + P^2/9 + Q^2/4\,.
\end{align}
\end{subequations}

At this point, use a Schwinger parametrisation to reexpress the product of denominators:
\begin{align}
\frac{1}{[D_{12} D_{3}]^2}
& = \int_0^\infty \!\! d\tau_1 \int_0^\infty \!\! d\tau_2 \, \tau_1 \tau_2 \,
{\rm e}^{-\tau_1 D_{12} - \tau_2 D_{3}}\,.
\end{align}
Then make a shift of the $p$ integration variable so that the exponent is an even function of both $p$, $q$, \emph{viz}.\
\begin{equation}
p\rightarrow p -\frac{\tfrac{4}{3}\tau_1 {\mathpzc  a}_1({\mathpzc a}_1-1)P +\tau_2 (1-2{\mathpzc a}_2)Q+\tfrac{2}{3}\tau_2P}
{2 ( \tau_1 {\mathpzc a}_1 (1-\mathpzc a_1) +\tau_2)}\,.
\end{equation}

Following these steps, one arrives at an expression of the following form:
\begin{align}
  \mathpzc T(Q;&P)  = \int_{\sf dp dq}
  \int_{d\mathpzc a_1 d\mathpzc a_2}
  \int_{d\tau_1 d\tau_2}\!\!\!\!  \tau_1 \tau_2 \nonumber \\
  &  {\cal N}^\prime(\mathpzc a_1, \mathpzc a_2,\tau_1, \tau_2,p,q,P,Q)
  {\rm e}^{-\tau_1 D_{12} - \tau_2 D_{3}}\,,
\end{align}
where the exponent also depends on all arguments, \emph{i.e}., $\mathpzc a_1, \mathpzc a_2,\tau_1, \tau_2,p,q,P,Q$, and is even in $p, q$.  At this point, one implements the SCI regularisation, \emph{viz}.\
\begin{align}
 & \mathpzc T(Q;P)  \to  \mathpzc T^{\rm ir}(Q;P) \\
&  = \int_{\tau_{\rm uv}}^{\tau_{\rm ir}} d\tau_1 d\tau_2\,  \tau_1 \tau_2
    \int_{d\mathpzc a_1 d\mathpzc a_2}\int_{\sf dp dq}
  \nonumber \\
  & \quad \times  {\cal N}^\prime(\mathpzc a_1, \mathpzc a_2,\tau_1, \tau_2,p,q,P,Q)
  {\rm e}^{-\tau_1 D_{12} - \tau_2 D_{3}}\,.
\end{align}

It is expressions of this type that we evaluate in order to deliver results for nucleon elastic form factors.

%%\bibliographystyle{elsarticle-num-names}
%%\bibliography{../../../CollectedBiB}

\end{document}